\documentclass[a4paper,11pt]{article}
\pdfoutput=1 % if your are submitting a pdflatex (i.e. if you have
\usepackage{jheppub} % for details on the use of the package, please
\usepackage{bm,amssymb,slashed,graphicx,multirow,soul,mathtools,xspace,array}

\usepackage{subfigure}
\usepackage{slashed}

\newcommand{\todo}[1]{{\color{red} \ifmmode\else[todo]\fi #1}}
\usepackage[usenames,dvipsnames]{xcolor}
     \definecolor{hgreen}{rgb}{0,.3,0}
     \definecolor{hred}{rgb}{.3,0,0}
     \definecolor{hblue}{rgb}{0,0,.3}
     \definecolor{LightGray}{gray}{0.95}

\newcommand{\dds}{D^{(*)}}

\newcommand{\cbar}{\bar{c}}
\newcommand{\Bbar}{\,\overline{\!B}{}}

\newcommand{\leff}{\ensuremath{\Lambda_{\rm eff}}\xspace}

\newcommand{\nt}{{\nu_\tau}}
\newcommand{\ntb}{{\bar\nu_\tau}}

\newcommand{\bddstn}{{\Bbar \to D^{(*)}\tau \bar\nu}}

\newcommand{\bddsln}{{\Bbar \to D^{(*)} l \bar\nu}}

\newcommand{\bddstaun}{{\Bbar \to D^{(*)} \tau \bar\nu}}
\newcommand{\bddstnR}{{\Bbar \to D^{(*)}\tau \bar N_R}}

\newcommand{\nn}{\nonumber}
\newcommand{\TeV}{\text{TeV}}
\newcommand{\GeV}{\text{GeV}}

\newcommand{\beq}{\begin{equation} }
\newcommand{\eeq}{\end{equation}} 
\newcommand{\bi}{\begin{itemize} }
\newcommand{\ei}{\end{itemize} }
\newcommand{\bea}{\begin{eqnarray}}
\newcommand{\eea}{\end{eqnarray}}

\newcommand{\Tr}{\mbox{Tr}\,}
\newcommand{\da}{\dagger}
\newcommand{\pda}{{\phantom{\da}}}
\newcommand{\al}{\alpha}

\newcommand{\hal}{{\hat{\alpha}}}

\definecolor{Red}{rgb}{1.,0.,0.}
\definecolor{Grn}{rgb}{0.,0.75,0.}
\definecolor{Blu}{rgb}{0.,0.,1.}

\def\vev#1{\left\langle #1\right\rangle}

\allowdisplaybreaks  

\graphicspath{{./figs/}}

\title{\boldmath $R(D^{(*)})$ from $W'$ and right-handed neutrinos}

\author[1,4]{Admir Greljo,}
\author[2]{Dean J. Robinson,}
\author[2,3]{Bibhushan Shakya,}
\author[2]{Jure Zupan}

\affiliation[1]{PRISMA Cluster of Excellence and Mainz Institute for Theoretical Physics, Johannes Gutenberg-Universit\"at Mainz, 55099 Mainz, Germany}
\affiliation[2]{Department of Physics, University of Cincinnati, Cincinnati, Ohio 45221,USA}
\affiliation[3]{Leinweber Center for Theoretical Phyics, University of Michigan, Ann Arbor, MI 48109, USA}
\affiliation[4]{Faculty of Science, University of Sarajevo, Zmaja od Bosne 33-35, 71000 Sarajevo, Bosnia and Herzegovina}

\emailAdd{admgrelj@uni-mainz.de}
\emailAdd{dean.robinson@uc.edu}
\emailAdd{bshakya@umich.edu}
\emailAdd{zupanje@ucmail.uc.edu}

\abstract{We provide an ultraviolet (UV) complete model for the $R(D^{(*)})$ anomalies, in which the additional contribution to 
semi-tauonic $b \to c$ transitions arises from decay to a right-handed sterile neutrino via exchange of a TeV-scale $SU(2)_L$ singlet $W'$. The model is based on an extension of the Standard Model (SM) 
hypercharge group, $U(1)_Y$, to the $SU(2)_V\times U(1)'$ gauge group, containing several pairs of heavy vector-like fermions. We present a comprehensive phenomenological survey of the model, 
ranging from the low-energy flavor physics, direct searches at the LHC, to neutrino physics and cosmology. We show that, while the $W'$ and $Z'$-induced constraints are important, 
it is possible to find parameter space naturally consistent with all the available data.
The sterile neutrino sector also offers rich phenomenology, including possibilities for measurable dark radiation, gamma ray signals, and displaced decays at colliders.
}

\preprint{LCTP-18-11, MITP/18-028}

\begin{document} 

\maketitle

%\flushbottom
\clearpage

\section{Introduction} 
Measurements of 
$|V_{cb}|$-independent ratios
\begin{equation}
 	R(\dds) = \frac{\mathcal{B}(\bddstn)}{\mathcal{B}(\bddsln)}\,, \qquad l = \mu, \,e\,,
\end{equation}
have been performed by the Babar~\cite{Lees:2012xj, Lees:2013uzd}, Belle~\cite{Huschle:2015rga, Abdesselam:2016cgx, Abdesselam:2016xqt}, and LHCb~\cite{Aaij:2015yra} collaborations. The results
exhibit a tension with the Standard Model (SM) expectations at the $4\sigma$ level when data from both $D$ and $D^*$ measurements are combined~\cite{HFAG} (see also Refs.~\cite{Bernlochner:2017jka,Bigi:2017jbd}).

The $b\to c\tau\ntb$ decays occur at tree-level in the SM. New Physics (NP) explanations of the $R(\dds)$ anomaly are therefore nontrivial, since they require new states close to the TeV scale. 
The NP contributions could, in principle, be due to a tree level exchange of a new charged scalar (see, e.g., \cite{Crivellin:2012ye,Celis:2012dk,Crivellin:2013wna}), 
a heavy charged vector (see, e.g., \cite{Greljo:2015mma,Boucenna:2016qad}), $W'$, or due to an exchange of a leptoquark, either vector or scalar (see, e.g., Refs.~\cite{Dorsner:2016wpm,Bauer:2015knc,Fajfer:2015ycq,Barbieri:2015yvd,Becirevic:2016yqi,Hiller:2016kry,Crivellin:2017zlb}). In all these cases, with the exception of Ref.~\cite{Becirevic:2016yqi}, the NP states couple to the SM neutrino. 
After the NP states are integrated out, they lead to four-fermion operators of the form $(\cbar \Gamma b)(\bar\ell \Gamma' \nt)$, where $\Gamma^{(\prime)}$ is an appropriate Dirac structure and $\nt$ is the active $\tau$ neutrino.

All these simplified models may naively produce $R(\dds)$ in approximate agreement with experiment (see, e.g.,~\cite{Fajfer:2012jt,Freytsis:2015qca,Bardhan:2016uhr}). 
They do, however, face a number of stringent constraints from complementary measurements. For instance, the pseudoscalar currents lead to too great a contribution to the $B_c$ lifetime
 from the enhanced $B_c\to \tau \ntb$ decay \cite{Li:2016vvp,Alonso:2016oyd,Celis:2016azn}, while the scalar currents are in tension with the $\bddstn$ differential rates 
 \cite{Lees:2013uzd,Huschle:2015rga,Freytsis:2015qca}. 
NP in the $b\to c\tau \ntb$ charged current transition necessarily implies a corresponding effect in the neutral currents, since $\nu_\tau$ is part of an $SU(2)_L$ doublet. 
Ref.~\cite{Faroughy:2016osc} used this observation to show that, both within effective field theory (EFT) and for simplified, UV complete, models,  the high-$p_T$ measurements of $p p \to \tau^+ \tau^-$ 
at the LHC already set important constraints on the NP explanations of the $R(\dds)$ anomaly. In addition, the leading one-loop electroweak corrections may lead to dangerously large 
contributions to the precisely measured $Z$ and $\tau$ decays \cite{Feruglio:2016gvd,Feruglio:2017rjo}. Generically, the models that avoid the $Z$ and $\tau$ decay contsraints lead to increased flavour 
changing neutral currents (FCNCs) in the down quark sector, e.g., in $B_s - \bar B_s$ mixing, $B\to K^{(*)} \nu_\tau \bar \nu_\tau$, and other observables. However, as shown in Ref.~\cite{Buttazzo:2017ixm}, 
it is still possible to simultaneously satisfy high-$p_T$ $\tau^+\tau^-$ production and electroweak precision observables, as well as the bounds on the FCNCs in the down quark sector, 
if the new dynamics is predominantly coupling to 
left-handed quarks and leptons with a very specific flavour structure. Representative UV models include: (i) a vector leptoquark singlets, $U_1^\mu$, that induces $B_s - \bar B_s$ mixing and 
$B\to K^{(*)} \nu_\tau \bar \nu_\tau$ only at one-loop~\cite{Buttazzo:2017ixm,DiLuzio:2017vat,Bordone:2017bld,Barbieri:2017tuq,Blanke:2018sro,Greljo:2018tuh}, or (ii) a pair of scalar leptoquarks, 
$S_1$ and $S_3$, with canceling tree-level contributions to $B\to K^{(*)} \nu_\tau \bar \nu_\tau$~\cite{Crivellin:2017zlb,Buttazzo:2017ixm,Marzocca:2018wcf}.

Here we follow an alternative approach: If the NP states couple instead to a \emph{right-handed} neutrino, many of the above constraints are avoided or suppressed.
That is, we examine the case that the $R(\dds)$ anomaly is generated by the $b \to c \tau \bar N_R$ transition, where $N_R$ is a light right-handed neutrino 
(in the remainder of the paper we denote $\nu$ = $N_R$ or $\nt$). We require $N_R$ to be light -- with mass $\lesssim \mathcal{O}(100)$\,MeV 
-- such that the measured missing invariant mass spectrum in the full $\bddstaun$ decay chain is not disrupted. (Whether heavier sterile neutrinos can be compatible with the data requires a full forward-folded study by the experimental collaborations.)

There are five possible four-Fermi operators involving such an additional, 
SM sterile, state  $N_R$ \cite{Robinson:2018EFT}, for earlier partial studies see \cite{Fajfer:2012jt,Becirevic:2016yqi,Cvetic:2017gkt}.
 Here, we focus on the specific case 
of an $SU(2)_L$ singlet $W'$-type mediator, which needs only carry a nonzero hypercharge. (This is in contrast to Ref.~\cite{Becirevic:2016yqi} which focused on the colored leptoquark mediator, that is more easily accessible in the direct searches at the LHC.) As such, the $W'$ may obtain its mass from the spontaneous breaking of an exotic non-Abelian symmetry. 
We show that the simplified model with $W'$ as a mediator may be UV completed within the so-called `$3221$' gauge model, and examine the relevant flavor, collider and cosmological constraints. 
Such a UV completion is rather minimal in its NP field content and can naturally lead to the largest NP effects in the $b \to c \tau\bar\nu$ transitions.

Further advantages of an $SU(2)_L$ singlet $W'$ interaction can be understood by comparision to, e.g., the $W'$ model of Ref.~\cite{Greljo:2015mma} (see also Ref.~\cite{Boucenna:2016qad}), 
which requires the $W'$ to be part of an $SU(2)_L$ triplet vector with the nearly degenerate $Z'$, as dictated by the $Z$-pole observables. Gauge invariance further requires that the flavor 
structures of $W'$ and $Z'$ couplings are related through the SM CKM mixing matrix. In our `$3221$' model, by contrast, these requirements are lifted, so that the observable effects of $Z'$ 
can be suppressed below the present experimental sensitivity, while at the same time one can still explain the $R(\dds)$ anomaly through the tree level exchange of the $W'$.

The paper is structured as follows. In Sec.~\ref{sec:EFT} we first present an EFT analysis of the $W'$ interaction with respect to the $\bddstaun$ data, followed by presentation of the UV complete 3221 model in Sec.~\ref{sec:3221}. Relevant collider and flavor constraints are explored in Sec.~\ref{sec:constr}. Verification of compatibility of the right-handed neutrino with neutrino phenomenology, including its cosmological history, is presented in Sec.~\ref{sec:neutrino}.
Section \ref{sec:conclusions} contains our conclusions. In App. \ref{app:FL} we describe the details on the flavor locking mechanism, while in App. \ref{sec:beyondminimal} we discuss the phenomenological implications, if the symmetry breaking of the new gauge interactions is non-minimal.

\section{The EFT  analysis}
\label{sec:EFT}
\subsection{Operators and effective scale}
We assume the SM field content is supplemented by a single new state, the right-handed sterile neutrino transforming as $N_R \sim (\bm{1}, \bm{1},0)$ under $SU(3)_c \times SU(2)_L \times U(1)_Y$. This state may couple to the SM quarks via any of the four dimension-6 operators 
\begin{align}
\label{eq:QS}
Q_{\text{SR}}&=\epsilon_{ab}\big(\bar q_L^a  d_R\big)\big(\bar \ell_L^b N_R \big),  & Q_{\text{SL}}&=\big(\bar u_R  q_L^a \big) \big(\bar \ell_L^a  N_R \big), \nn
\\ Q_{\text{T}}&=\epsilon_{ab} \big(\bar q_L^a \sigma^{\mu\nu} d_R\big)\big(\bar \ell_L^b\sigma_{\mu\nu} N_R \big), & Q_{\text{VR}}&=\big(\bar u_R \gamma^\mu  d_R\big)\big(\bar e_R \gamma_\mu N_R \big),
\end{align}
suppressing for now the generational indices (an operator containing a vector current with left-handed quarks is also possible, but requires two Higgs insertions and is thus dimension-8). 
We focus on the operator $Q_{\rm VR}$. This is generated in a simplified model by a tree level exchange of the  $W' \sim (\bm{1},\bm{1}, +1)$ mediator, with the interaction Lagrangian,
\begin{equation}\label{eq:LW'}
	\mathcal{L}_{W'} =  \frac{g_V}{\sqrt{2}}  c_q^{ij} \bar u_R^i\slashed W' \negmedspace d_R^j  + \frac{g_V}{\sqrt{2}} c_N^{i}  \bar N_R \slashed W' \negmedspace e_R^i +{\rm h.c.}\,,
\end{equation}
with $g_V$ an overall coupling constant, while $c_q^{ij}, c_N^i$ coefficients encode the flavor dependence of $W'$ interactions.
Restoring the flavor structure to $Q_{\text{VR}}$, the $b \to c \ell N_R$ decay then arises from
\begin{equation}
	\label{eqn:QVRL}
	\mathcal{L}_{\text{VR}} =  \frac{C_{ij,k}}{\leff^{2}}\big(\bar u_R^i \gamma^\mu  d_R^j\big)\big(\bar e_R^k \gamma_\mu N_R \big)\,, \qquad C_{ij,k} = \frac{g_V^2 c_q^{ij} c_N^k\leff^2}{2 m_{W'}^2}\,,
\end{equation}
with $i,j,k=1,\ldots, 3$ the generation indices. Above we have defined an effective scale
\begin{equation}
\label{eq:Lambda:eff}
	\leff=\big(2\sqrt 2 G_F V_{cb}\big)^{-1/2}\simeq 0.87 \, \left(\frac{40 \times 10^{-3}}{V_{cb}}\right)^{1/2}\,\TeV\,,
\end{equation}
choosing a phase convention in which $V_{cb}$ is real. We work in the mass basis, such that setting $i=2, j=3, k=3$ in Eq.~\eqref{eqn:QVRL} generates the operator $\big(\bar c_R \gamma^\mu  b_R\big)\big(\bar \tau_R \gamma_\mu N_R \big)$.  The definition for $\leff$ in \eqref{eq:Lambda:eff} is chosen such that the rate for the $\bddstnR$ decay is normalized to the SM rate for the $\bddstn$ process at $C_{23,3}=1$. The $\bddstn$ decays become an incoherent sum of two contributions: from the SM decay, $b\to c\tau \ntb$, as well as from the new decay channel, $b\to c\tau \bar N_R$. The NP contributions therefore necessarily increase both of the $\bddstn$ branching ratios above the SM expectation, in agreement with the direction of the experimental observations for $R(\dds)$. 

\begin{figure}[t]
\begin{center}
\includegraphics[width=8.5cm]{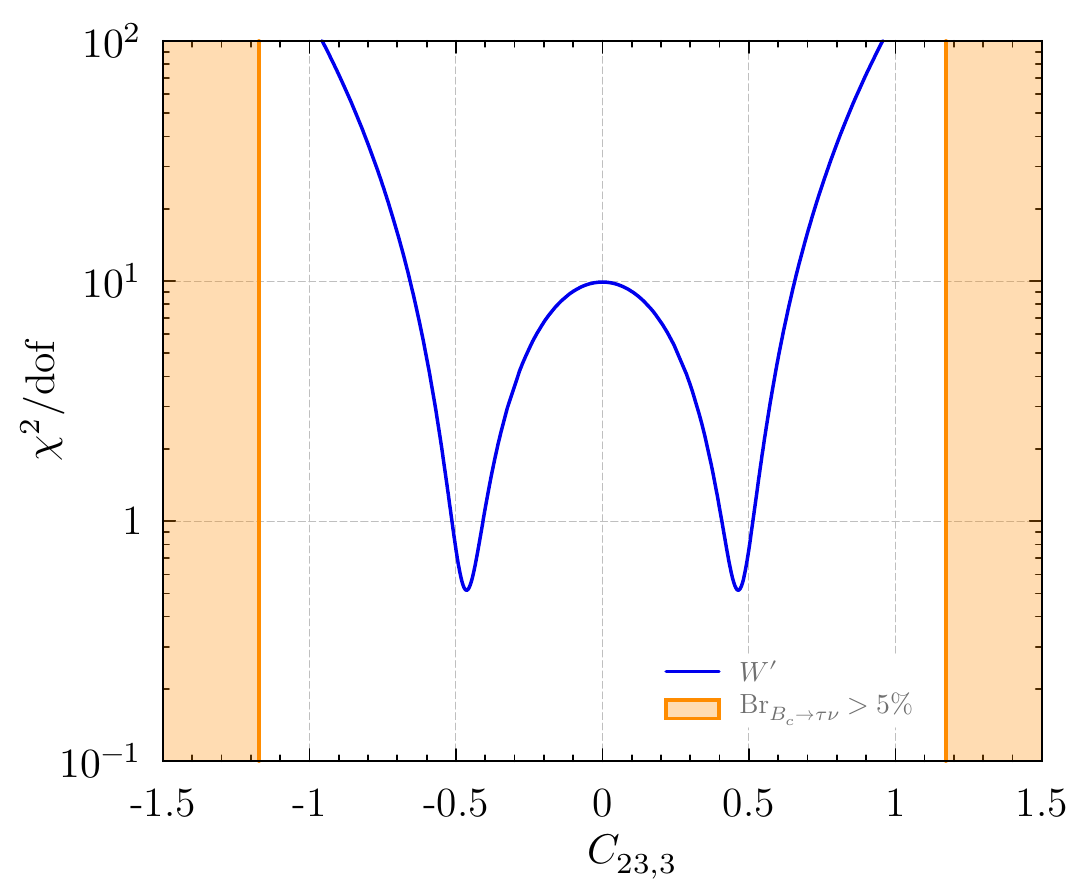}
\end{center}
\caption{The $\chi^2/\text{dof}$ distribution (blue) for the fit of the $R(\dds)$ predictions in the $Q_{\text{VR}}$ effective theory, Eq. \eqref{eqn:QVRL}, to the current world average~\cite{HFAG}. Also shown (shaded orange) are exclusion regions for $\text{Br}(B_c \to \tau \nu) \gtrsim 5\%$.} 
\label{fig:fits}
\end{figure}

\subsection{Fit to the $\bddstaun$ data}
The addition of $b\to c\tau \bar N_R$ transitions to the SM $b\to c\tau \ntb$ process does not change significantly the differential distributions (see Fig.~\ref{fig:1Dhistos} and discussion in Sec.~\ref{sec:eftdd} below). Computation of the $\Bbar \to (D^* \to D Y)(\tau \to \nu X) \bar \nu$ differential distributions and corresponding $R(\dds)$ predictions are obtained from the expressions in Ref.~\cite{Ligeti:2016npd}, making use of the form factor fit `$L_{w\ge1}\text{+SR}$' of Ref.~\cite{Bernlochner:2017jka}. This fit was  performed at next-to-leading order in the heavy quark expansion, utilizing the recently published unfolded Belle $\bddsln$ data~\cite{Abdesselam:2017kjf} and state-of-the-art lattice calculations beyond zero recoil~\cite{Bailey:2014tva, Lattice:2015rga}.  Fitting the $R(\dds)$ predictions to the current experimental world averages~\cite{HFAG},
\begin{equation}
	\label{eqn:RDDsdata}
	R(D) = 0.407 \pm 0.046\,, \qquad R(D^*) = 0.304\pm 0.015\,, \qquad \text{corr.} = -0.20\,,
\end{equation}
gives the $\chi^2/\text{dof}$ 
as a function of $C_{23,3}$  shown in Fig.~\ref{fig:fits} ($\text{dof} = 2$). The best fit value is obtained for $C_{23,3} \simeq 0.46$, with $\chi^2/\text{dof} \simeq 0.5$, to be compared with $\chi^2/\text{dof} \simeq 10.$ at the SM point, $C_{23,3}=0$. This best fit corresponds to
\begin{equation}
 	\leff/\sqrt{C_{23,3}} \simeq \,1.3\,\bigg[\frac{40 \times 10^{-3}}{V_{cb}}\bigg]^{1/2}\,\TeV\,,
\end{equation}
and in the $W'$ simplified model to the $W'$ mass of
\begin{equation}
	m_{W'} \simeq 540  \big[c_q^{23}c_N^{3}\big]^{1/2} \bigg[\frac{g_V}{0.6}\bigg] \bigg[\frac{40 \times 10^{-3}}{V_{cb}}\bigg]^{1/2}\,\GeV\,, \label{eq:fit}
\end{equation}
in which we normalized,  for illustration, $g_V$ to the approximate value of the SM weak coupling constant, $g_2$. 

The additional $W'$ current also incoherently modifies the $B_c \to \tau \bar\nu$ decay rate with respect to the SM contribution, such that
\begin{equation}
\label{eq:Bctaunu}
	\text{Br}(B_c \to \tau \bar\nu) = \frac{\tau_{B_c} f_{B_c}^2m_{B_c} m_\tau^2}{64 \pi \leff^4}\big(1 - m_\tau^2/m_{B_c}^2\big)^2\Big[1 + |C_{23,3}|^2\Big]\,,
\end{equation}
with $f_{B_c} \simeq 0.43$\,GeV~\cite{Colquhoun:2015oha} and $\tau_{B_c} \simeq 0.507$\,ps~\cite{PDG}. Conservatively we require $\text{Br}(B_c \to \tau\bar\nu) < 5\%$~\cite{Li:2016vvp,Alonso:2016oyd}. In Fig.~\ref{fig:fits} we show the corresponding exclusion region for $|C_{23,3}|$ (orange shaded regions), which is far from the best fit region.

\begin{figure}[t]
\centering{
\includegraphics[width = 0.47\linewidth]{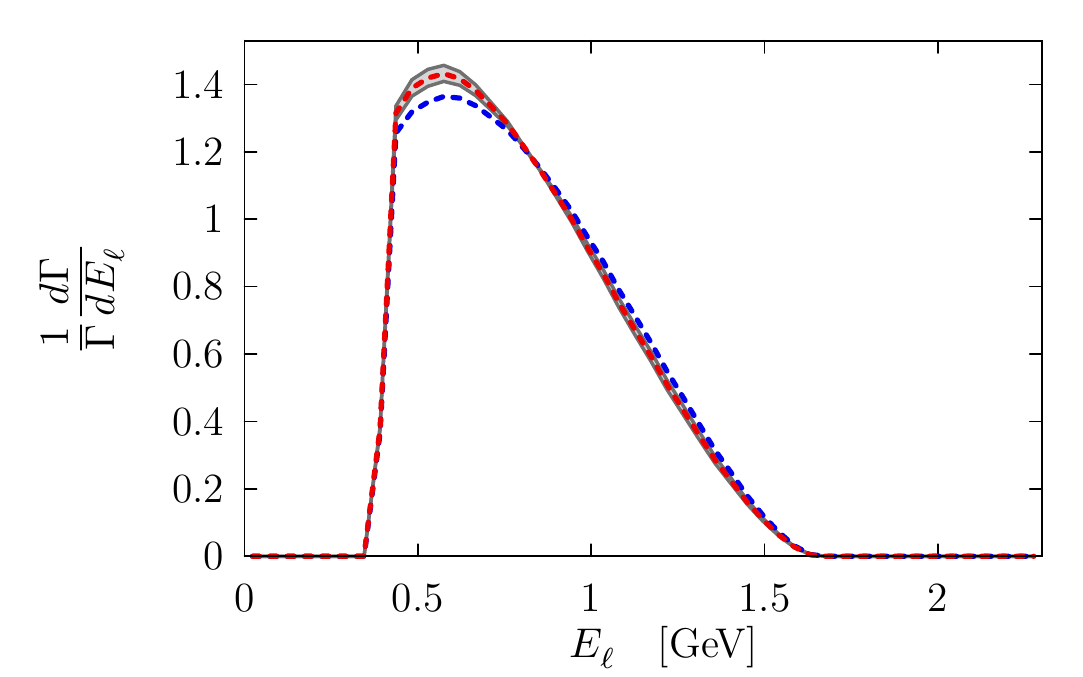}\hfill
\includegraphics[width = 0.47\linewidth]{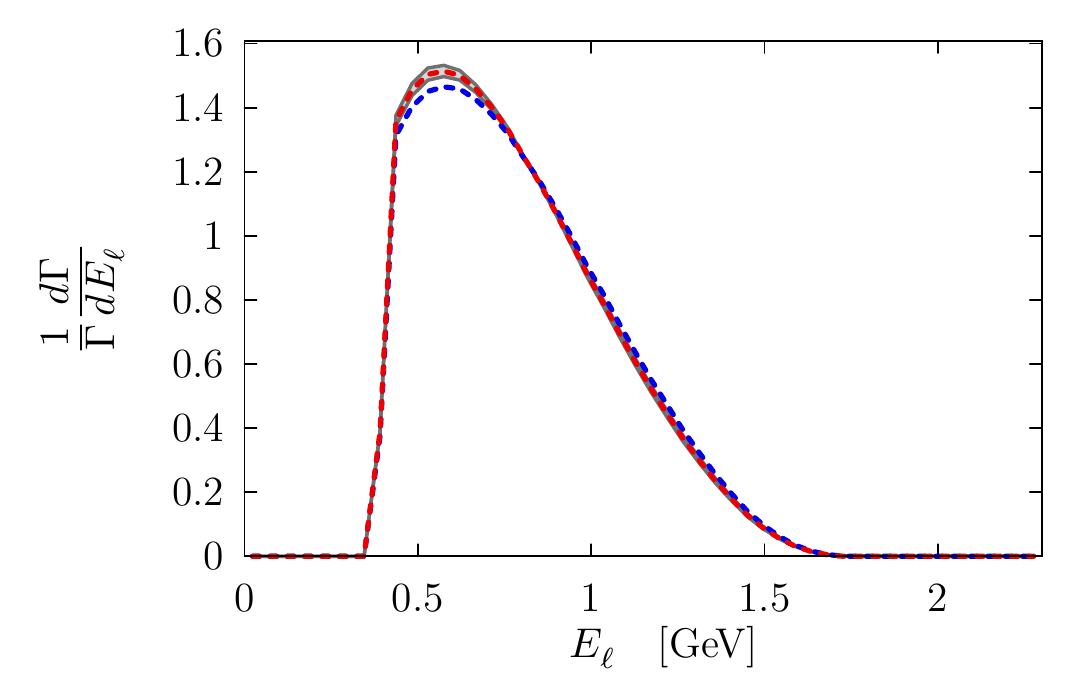} \\
\includegraphics[width = 0.47\linewidth]{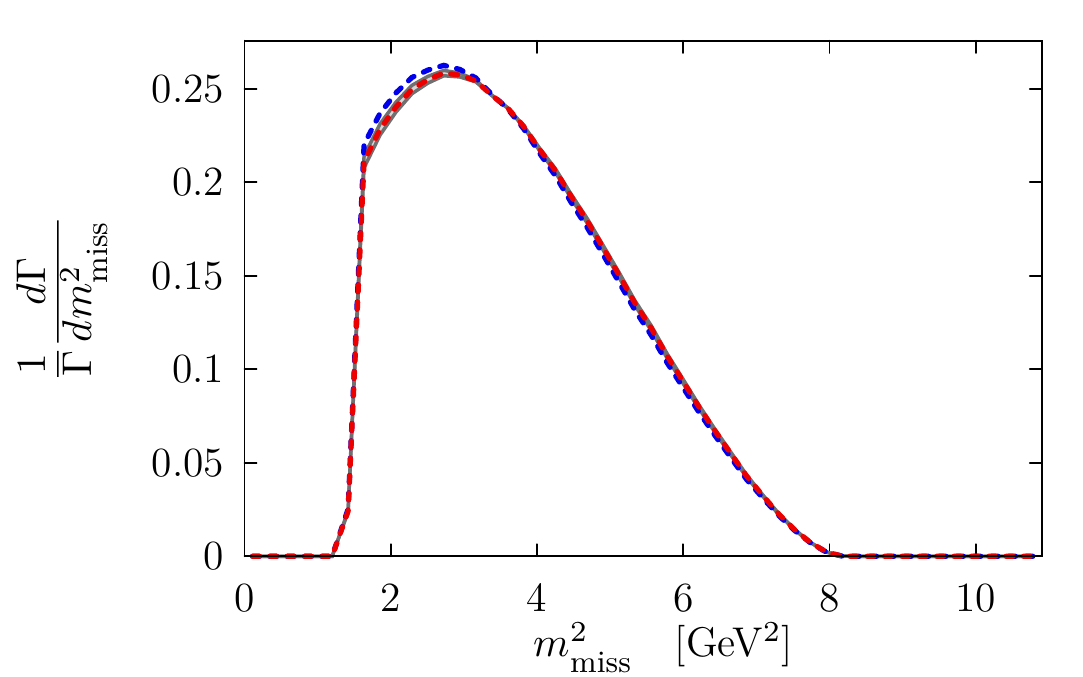} \hfill 
\includegraphics[width = 0.47\linewidth]{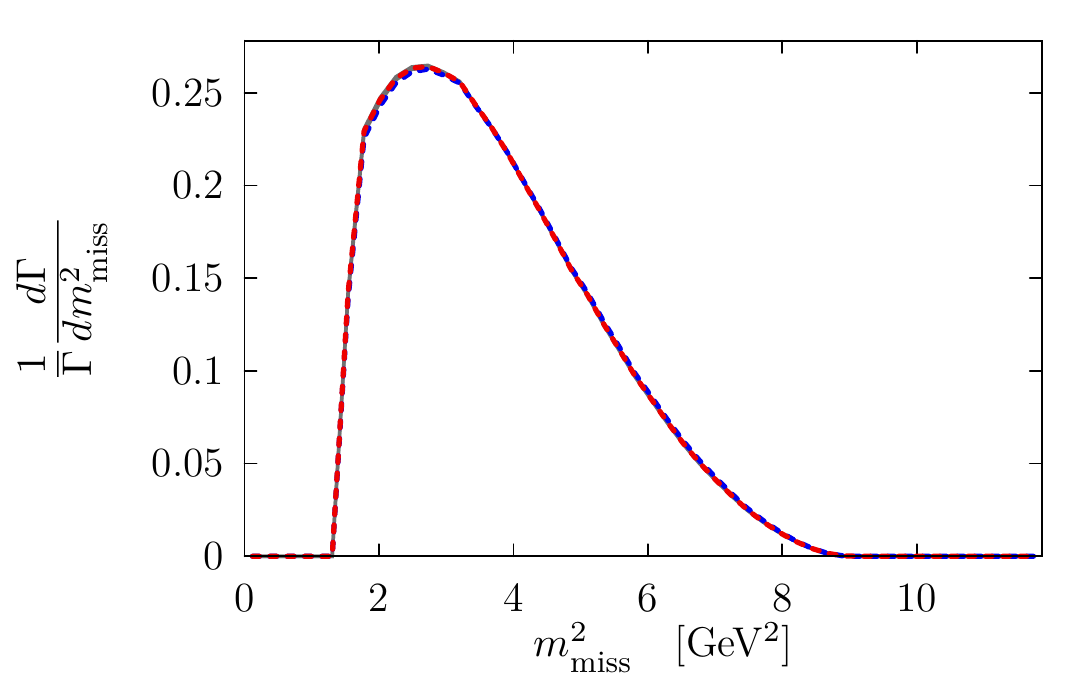}\\
\includegraphics[width = 0.47\linewidth]{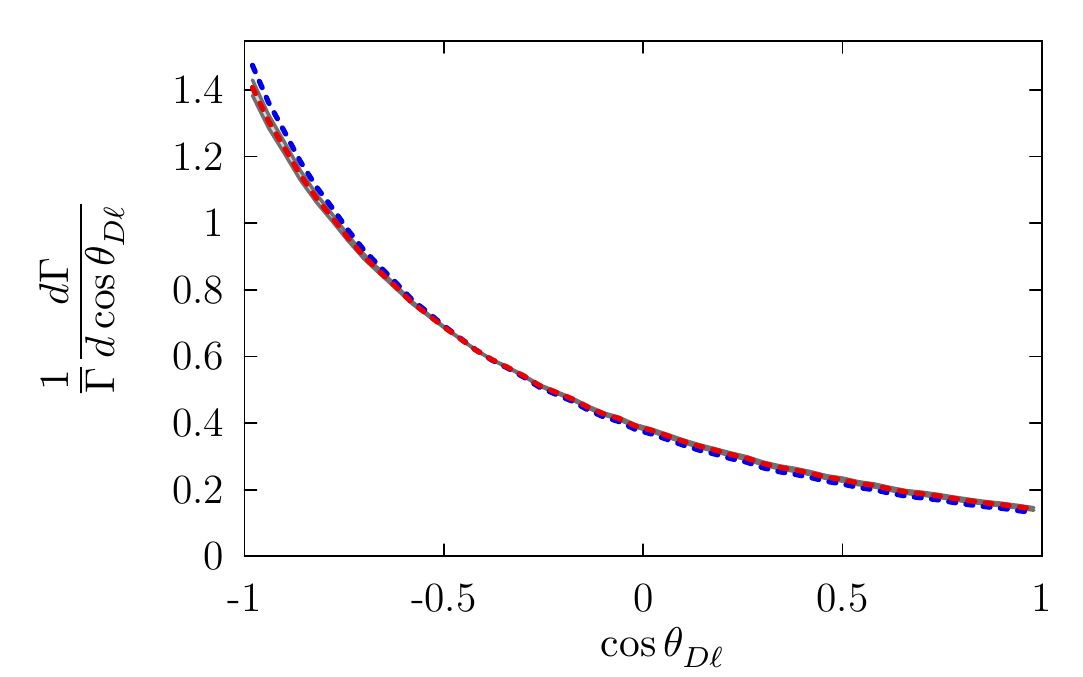} \hfill
\includegraphics[width = 0.47\linewidth]{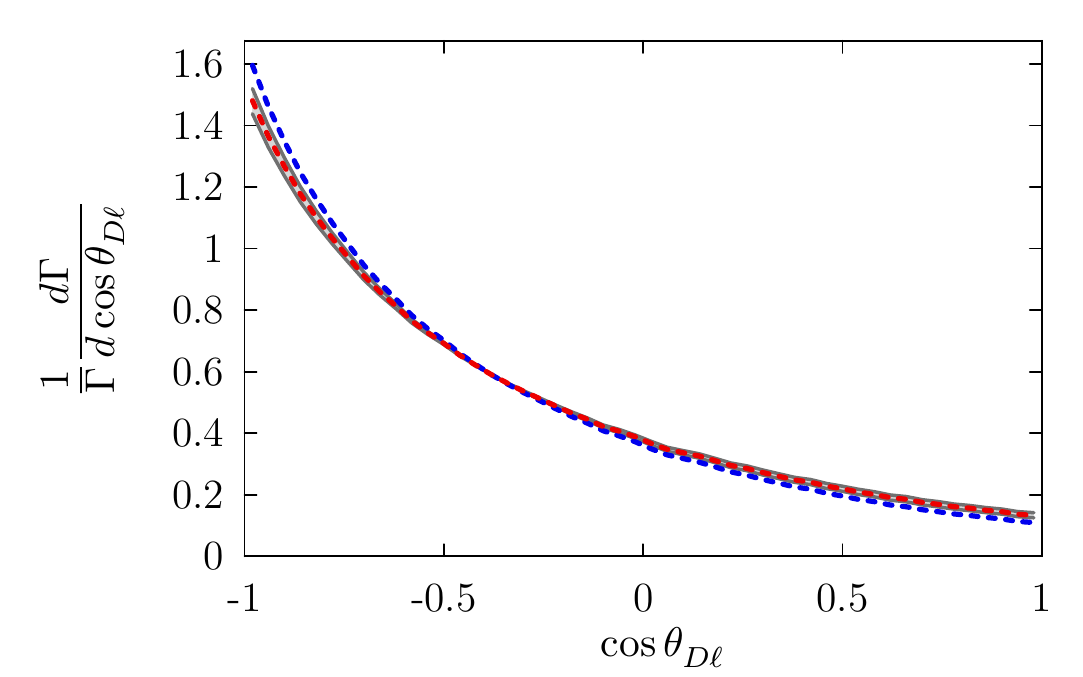} \\
}
\caption{Kinematic distributions in the $B$ rest frame for couplings ranging over $C_{23,3} \in [0.26, 0.66]$ (gray regions) with phase space cuts~\eqref{eqn:PScut}, for $\Bbar \to (D^* \to D\pi)(\tau \to \ell \bar\nu_\ell \nt)\bar\nu$ (left) and $\Bbar \to D(\tau \to \ell \bar\nu_\ell \nt)\bar\nu$ (right). The blue (red) dashed curves show the SM (SM+$W'$ best fit, $C_{23,3} = 0.46$, massless $N_R$).}
\label{fig:1Dhistos}
\end{figure}

\subsection{Differential distributions}
\label{sec:eftdd}
Crucial to the reliability of the above fit results is the underlying assumption that the differential distributions, and hence experimental acceptances, of the $\bddstn$ decays are not significantly modified in the presence of the $W'$ current. Experimental extraction of $R(\dds)$ relies on a simultaneous float of background and signal data, and can be significantly model dependent (cf., e.g., the measured values of $R(\dds)$ for the SM versus Type II 2HDM in Ref.~\cite{Huschle:2015rga}). In Fig~\ref{fig:1Dhistos} we show normalized differential distributions for the detector observables $E_\ell$, $m^2_{\text{miss}}$ and $\cos\theta_{D\ell}$ arising from the cascades $\Bbar \to (D^* \to D\pi)(\tau \to \ell \bar\nu_\ell \nt)\bar\nu$ and $\Bbar \to D(\tau \to \ell \nu \nu)\nu$, for the SM versus SM+$W'$ theories, taking $N_R$ to be massless, and applying the phase space cuts,
\begin{equation}
	\label{eqn:PScut}
	q^2 = (p_B - p_{\dds})^2 > 4~\text{GeV}^2\,, \qquad E_{\ell} > 400~\text{MeV}\,,\qquad m^2_{\text{miss}} > 1.5~\text{GeV}^2\,,
\end{equation}
as an approximate simulation of the measurements performed in Refs.~\cite{Lees:2013uzd,Huschle:2015rga}. These distributions are generated as in Ref.~\cite{Ligeti:2016npd}, using a preliminary version of the \texttt{Hammer} library~\cite{Hammer_paper}. In each plot, we show the variation in shape over the range $C_{23,3} = 0.46 \pm 0.2$, corresponding to a range greater than the $99\%$CL, and for the SM ($C_{23,3} = 0$). One sees that the variation in shape is small over this range. The variation in other observables, such as $q^2$, is not shown, since it is even smaller. This gives us good confidence that the measured $R(\dds)$ in Eq.~\eqref{eqn:RDDsdata} well-approximate the values that would be measured for a SM+$W'$ model template.

%%%%%%%%%%%%%%%%%%%%%%%%%%%
\section{Explicit UV completion: The `$3221$' gauge model}
%%%%%%%%%%%%%%%%%%%%%%%%%%%
\label{sec:3221}
A massive vector requires a UV completion. We consider a `3221'-type gauge theory, with a gauge group $\mathcal{G}  = SU(3)_c \times SU(2)_L \times SU(2)_V \times U(1)'$. 
The $U(1)'$ together with the $SU(2)_V$ symmetry will generate heavy vectors under spontaneous symmetry breaking $SU(2)_V \times U(1)' \to U(1)_Y$.
Our notation for the gauge fields in the ${\cal G}$-symmetric phase is $G^a_\mu$, $W^i_\mu$, $W'^j_\mu$, and $B'_\mu$, respectively, with $g_s$, $g_L$, $g_V$, and $g'$ the corresponding gauge couplings.  
The content of the model is shown in Table~\ref{tab:fieldcontent}: Three generations of SM-like chiral field content, denoted by primes, is extended by a right-handed neutrino $\nu'_R$. Also included are one or more generations of vector-like quarks and leptons, $Q_{L,R}'^i$ and $L_{L,R}'^i$ that transform as doublets under $SU(2)_V$.  We will consider the phenomenological implications for the cases where either one, two, or three sets of vector-like fermions are introduced. In the remainder of this section, we give a detailed account of this UV completion, while the related phenomenology is discussed in Section~\ref{sec:constr}.

\begin{table}[t]
\renewcommand{\arraystretch}{1.1}
\begin{center}
\begin{tabular}{ccccc}
\hline \hline
 {\rm Field } & $SU(3)_c$    &   $SU(2)_L$   & $SU(2)_{V}$ & $U(1)'$ \\ \hline
\multicolumn{5}{c}{ \rm SM-like chiral fermions} \\ %\hline
 $q_{L}'^{i}$     &  {\bf 3}  &  {\bf 2}  & {\bf 1} & 1/6       \\ 
 $\ell_{L}'^{i}$  &  {\bf 1}  &  {\bf 2}  &  {\bf 1} &-1/2   \\ 
 $u_{R}'^{i}$    &  {\bf 3}  &  {\bf 1}  & {\bf 1} & 2/3      \\ 
 $d_{R}'^{i}$    &  {\bf 3}  &  {\bf 1}  & {\bf 1} & -1/3      \\ 
 $e'^{i}_{R}$   &  {\bf 1}  &  {\bf 1}  &  {\bf 1} & -1     \\ 
 $\nu'^{i}_{R}$   &  {\bf 1}  &  {\bf 1}  &  {\bf 1} & 0     \\ 
 \hline
 \multicolumn{5}{c}{ \rm Extra vector-like fermions} \\ %\hline
$Q_{L,R}'^i$  & {\bf 3}  &   {\bf 1}  &  {\bf 2}  & 1/6      \\ 
$L_{L,R}'^i$  & {\bf 1}  &   {\bf 1}  &  {\bf 2}  & -1/2    \\ 
   \hline
 \multicolumn{5}{c}{ \rm Scalars} \\ %\hline
$H$  & {\bf 1}  &  {\bf 2}  &  {\bf 1}  & 1/2     \\ 
$H_{V}$  & {\bf 1}  &  {\bf 1}  &  {\bf 2}  & 1/2    \\ 
   \hline\hline
\end{tabular}
\end{center}
\caption{Matter content of the model in the unbroken phase of gauge group ${\cal G}$. The flavour index $i \in \{1,2,3\}$. Singlet representation is denoted with $ {\bf 1}$, while fundamental of $SU(3)$ ($SU(2)$) is ${\bf 3}$ (${\bf 2}$). The last column shows the $Y'$ quantum number.
\label{tab:fieldcontent}}
\end{table}

\subsection{Gauge symmetry and the spontaneous symmetry breaking pattern} 

The gauge group ${\cal G}$ is spontaneously broken in two steps, first ${\cal G} \to {\cal G}_{\textrm{SM}}\equiv SU(3)_c \times SU(2)_L \times U(1)_Y$, and then ${\cal G}_{\rm SM}\to U(1)_{\rm em}$. The first step of spontaneous symmetry breaking, ${\cal G} \to {\cal G}_{\textrm{SM}}$, occurs when the scalar, 
$H_V$ obtains a nonzero vacuum expectation value (vev),
\begin{equation}
\label{eq:HVvev}
\vev{H_V} = \frac{1}{\sqrt{2}}\begin{pmatrix} 0 \\ v_V \end{pmatrix}\,.
\end{equation}
 This results in three Goldstone modes being eaten by the $W'^{\pm}$ and $Z'$ gauge bosons, giving
\begin{align}
W'^{\pm}_\mu &  = \frac{1}{\sqrt{2}}(W'^{1}_\mu \mp i W'^{2}_\mu)~, & m_{W'} &= \frac{g_V v_V}{2}~,\\
Z'_ \mu & = \cos \theta_V W'^{3}_\mu - \sin \theta_V B'_\mu~, &  m_{Z'} &= \frac{m_{W'}}{\cos \theta_V}~,
\label{eq:masses}
\end{align}
where $\tan \theta_V = g' / g_V$. In the following, we will use the notation 
\beq
\label{eq:cV:notation}
c_V\equiv \cos\theta_V, s_V\equiv \sin\theta_V{\rm ~~ and~~}t_V\equiv \tan\theta_V.
\eeq 

The $H_V$ vev in Eq. \eqref{eq:HVvev} breaks $SU(2)_V\times U(1)'\to U(1)_Y$. The unbroken generator, $Y = T^3_V + Y'$, corresponds to the massless SM hypercharge gauge boson, $B_ \mu = s_V W'^{3}_\mu + c_V B'_\mu$. Here, $T^3_V$ is the diagonal generator of $SU(2)_V$. 
The hypercharge gauge coupling is
\beq
\label{eq:gY:gV}
g_Y = \frac{g_V g'}{\sqrt{g_V^2 + g'^2}}.
\eeq
This relation fixes the mixing angle to $t_V=g_Y/(g_V^2-g_Y^2)^{1/2}$, so that $g_V$ needs to be larger than $g_Y\simeq 0.36$. For large values of $g_V$ we have $t_V\simeq g_Y/g_V$.

The second step in spontaneous symmetry breaking is the usual electroweak symmetry breaking within the SM, due to the Higgs vev, $\langle H\rangle =(0,v_{\rm EW}/\sqrt{2})^T$, with the  SM Higgs spanning the $H \sim ({\bf 1},{\bf 2},{\bf 1},1/2)$ representation of ${\cal G}$.  For simplicity we assume here and in the remainder of the manuscript that the mixed quartic, $\mathcal{L} \supset - \lambda_{\textrm{mix}} (H^\dagger H) (H_V^\dagger H_V)$ is negligible. As a consequence, we can neglect the mixing between the two real scalar excitations, the SM Higgs, $h$, and the heavy Higgs, $h_V$, simplifying the discussion. 

The NP contributions to $R(\dds)$ scale as $g_V^4/m_{W'}^4\sim 1/v_V^4$.  In order to explain $R(\dds)$ the vev  $v_V$ should not be too large, while direct searches require $m_{W'}\sim g_V v_V$ to be large. A phenomenologically viable solution is obtained for $g_V\gtrsim 1$,  which then requires $g'\simeq g_Y \approx 0.36$ (for a detailed numerical analysis see Section \ref{sec:constr}).
In this case the $\theta_V$ mixing angle is small, $\sin\theta_V\lesssim 0.3$, and thus $Z'$ and $W'$ masses are almost degenerate $m_{Z'}-m_{W'}\lesssim 0.05 m_{W'}$ in the minimal ${\cal G} \to {\cal G}_{\textrm{SM}}$ breaking scenario. Because of stringent constraints from the $Z'$ searches, an extra source of mass splitting between the $Z'$ and $W'$ might be required in some cases: A possibility that we partially explore in Appendix \ref{sec:beyondminimal}.

\subsection{Matter content and new Yukawa interactions} 
In order to make the phenomenology of the model more tractable, and the notation more streamlined, we make the simplifying assumption that only one generation of the vector-like fermions, $Q_{L,R}'^i$ and $L_{L,R}'^i$, have appreciable couplings to the SM quarks and leptons. We denote the corresponding fields by just $Q_{L,R}'$, $L_{L,R}'$. They decompose under the SM gauge group as
\beq
Q'_{L,R} = \left(\begin{array}{c}
U'_{L,R}\\
D'_{L,R}
\end{array}\right)~, \; \; \; \;
L'_{L,R} = \left(\begin{array}{c}
N'_{L,R}\\
E'_{L,R}
\end{array}\right)~,
\eeq
where $U'_{L,R} \sim ({\bf 3},{\bf 1},2/3)$, $D'_{L,R}\sim({\bf 3},{\bf 1},-1/3)$, $N'_{L,R} \sim ({\bf 1},{\bf 1},0)$, and $E'_{L,R}\sim ({\bf 1},{\bf 1},0)$ under ${\cal G}_{\rm SM}$. This is the minimal field content required to generate the $b\to c\tau N_R$ transitions. An additional pair of vector-like fermions, $\tilde Q_{L,R}'$, $\tilde L_{L,R}'$, are assumed to have very small couplings to the SM fields and will thus be relevant only in the discussion of collider searches. 
 
 The mixing of the SM-like chiral fermions and the vector-like fermions, $Q_{L,R}'$, $L_{L,R}'$, occurs through the following Yukawa interactions in the Lagrangian,
\begin{align}
\label{eq:Yukawas}
\mathcal{L}_{\textrm{Yuk}} =  \mathcal{L}^{\textrm{SM}}_{\textrm{Yuk}} 
	& - \lambda_d^i \bar Q'_L H_V d'^i_R - \lambda_u^i \bar Q'_L \tilde H_V u'^i_R\\
	& - \lambda_e^i \bar L'_L H_V e'^i_R - \lambda_\nu^i \bar L'_L \tilde H_V \nu'^i_R + \textrm{h.c.}~,
\end{align}
where the Yukawa interactions between the SM fields are, as usual,
\begin{equation}
\label{eq:SMYukawa}
\mathcal{L}^{\textrm{SM}}_{\textrm{Yuk}} =   - \bar q'_L Y_d H d'_R - \bar q'_L Y_u \tilde H u'_R - \bar \ell'_L Y_e H e'_R - \bar \ell'_L Y_\nu \tilde H \nu'_R + \textrm{h.c.}~,\\
\end{equation}
and $\tilde H_{(V)} = \epsilon H_{(V)}^*$. The $Q'$ and $L'$ mass terms are $M_Q \bar Q'_L Q_R' - M_L \bar L'_L L_R'$. 
Without loss of generality we can take $M_{Q,L}$ to be real positive, and set, using the flavour group rotations, $Y_d = Y_d^{\rm{diag}}$, $Y_e = Y_e^{\rm{diag}}$, and 
$Y_u = V^\dagger Y_u^{\rm{diag}}$, with $V$ a unitary $3\times 3$ matrix, and $Y_{u,d,e}^{\rm diag}$ meaning a diagonal $3\times 3$ matrix with real positive entries. 
(The neutrino sector is discussed separately below, in Section \ref{sec:neutrinomasses:first}.) 
To simplify the discussion we take all the $\lambda_f^i$ to be real. In this basis,  the couplings  $\lambda_d^3 \equiv \lambda_b$, $\lambda_u^2 \equiv \lambda_c$, $\lambda_e^3 \equiv \lambda_\tau$, 
and $\lambda_\nu^3$, need to be large in order to explain the $R(\dds)$ anomaly. 

Before $H_V$ and $H$ obtain vevs, the vector-like fermions $Q_{L,R}', L_{L,R}'$ have masses $M_{Q,L}$, while the SM fermions are massless. The $H_V$ vev induces the mixing between the heavy vector-like fermions and the right-handed SM fermions. After the electroweak symmetry is broken by the Higgs vev, the SM fermions become massive, inducing mixing with the left-handed SM fermions.  We first investigate the mixing between the vector-like fermions and the SM fermions for the simplified case of a $2 \times 2$ system, taking as an illustration the limit of only the bottom quark, $b_R'$, coupling to the vectorlike fermion $Q_{L,R}'$.

 For $\langle H_V\rangle \ne 0$, but still keeping $\langle H\rangle= 0$, the mass eigenstates are the $D_R, b_R$ fermions with (left-handed components are not mixed so that $D_L=D_L'$) 
\beq
D_R = \cos\theta_{b_R} D'_R+\sin\theta_{b_R} b'_R, {\rm ~~and~~~}  b_R = -\sin\theta_{b_R} D'_R + \cos\theta_{b_R} b'_R,
\eeq
 where the mixing angle satisfies, 
 \beq
 \tan\theta_{b_R} = \frac{\lambda_b v_V}{ \sqrt{2} M_Q}.
 \eeq 
 The heavy quark $D$ has mass 
 \beq\label{eq:MD}
 M_D \equiv M_Q\sqrt{1 + \tan^2\theta_{b_R}}\,, 
 \eeq
 while $b_R$ remains massless. 
 
 After electroweak symmetry breaking due to the Higgs vev, $v_{\rm EW}\ne 0$, also the left-handed fields, i.e., the down component of $q_L'$, and the $D_L$ mix. The corresponding left-handed mixing angle is
 \beq\label{eq:sbL}
 \sin\theta_{b_L} \approx \frac{m_b}{m_D} \tan\theta_{b_R},
 \eeq
in which the mass of the light quark is  
 \beq
m_b \approx \frac{v_{\rm EW}}{\sqrt{2}} Y_d^{(33)} \cos\theta_{b_R},
\eeq
while the mass of the heavy state, $D$, remains  $\approx M_D$.

The above analysis extends straightforwardly to the three generations of SM quarks. In the first step now a linear combination of SM quarks mixes with $D_R$ when $H_V$ obtains a vev, $\langle H_V \rangle \ne 0$. The expressions for the second step, the electroweak symmetry breaking, can be found in Section IV of Ref.~\cite{Fajfer:2013wca}, where a general phenomenology of mixings with a singlet down-like vector-like quark has been worked out. 
The left-handed mixing \eqref{eq:sbL} results in a tree-level modification of the effective $Z$ boson couplings, which were precisely measured at LEP. In the limit of large $\tan\theta_{b_R}\gg1$, i.e., in the limit $\lambda_b v_V\gg M_Q$, the left-handed mixing is given by $\sin\theta_{b_L} \approx m_b / M_Q$, implying a lower limit $M_Q \gtrsim 100$~GeV~\cite{Fajfer:2013wca}.  Repeating the same analysis for charm, we find a comparable, yet somewhat less stringent, bound on $M_Q$. Similar bounds apply also on $M_L$ from $Z\to \tau^+ \tau^-$ and lepton flavor universality measurements in $\tau$ decays. 

As we will discuss later on, the explanation of $\mathcal{R}(\dds)$ anomaly requires $\sin\theta_{b_R}$, $\sin\theta_{c_R}$, and $\sin\theta_{\tau_R}$ to be ${\mathcal O}(1)$. The analysis above the implies that  one can take as a realistic benchmark $\tan\theta_{b_R,c_R, \tau_R}\approx 10$, i.e.,  the case where right-handed bottom and charm quarks, as well as the right-handed tau are mostly composed from the corresponding vector-like states, so that $\sin\theta_{b_R,c_R,\tau)R}\approx 1$, and $\cos\theta_{b_R,c_R,\tau_R}\approx 0.1$. 
We explore the phenomenology of the $\tan\theta_{b_R,c_R,\tau_R}\gg1$ limit ($\lambda_{b,c (\tau)} v_V\gg M_{Q(L)}$) in detail  in Section \ref{sec:constr}.

\subsection{Gauge boson interactions} 
For later convenience we also give the couplings of $W'^\pm$ and $Z'$ to fermions, all of which come from the covariant derivatives in the kinetic terms of the fermions. In the interaction basis we have,
\beq
\begin{split}
\label{eq:couplings}
\mathcal L &\supset \frac{g_V}{\sqrt{2}} \Big(\bar U' \gamma^\mu D' + \bar N' \gamma^\mu E'\Big) W'^+_\mu + \textrm{h.c.}\\
&+\frac{g_Y}{ s_V c_V} \sum_{F'} \big[ T_V^3(F')-s_V^2 ~ Y(F')\big] \,\big(\bar F' \gamma^\mu F' \big)Z'_\mu\\
&- g_Y t_V \sum_{f'} Y(f') \, \big(\bar f' \gamma^\mu f' \big)Z'_\mu~,
\end{split}
\eeq
where $T_V^3$ and $Y$ are the corresponding fermion quantum numbers under $SU(2)_V\times U(1)'$. The summation is over $F'=U_{L,R}'^i, D_{L,R}'^i, N_{L,R}'^i, E_{L,R}'^i$, and $f' = q_L'^i, \ell_L'^i, u_R'^i, d_R'^i, e_R'^i$. In the absence of fermion mass mixing the $W'$ only couples to the vector-like fermions. The $Z'$, however, also couples to the $f'$ fermions. A phenomenologically viable scenario requires $t_V \ll 1$ in order to suppress $p p \to Z'$ production from the valence quarks (see Fig.~\ref{fig:LHCZp} and discussion in Section~\ref{sec:constr}). 

We are now ready to map the above results to the notation we used for the EFT analysis of $R(\dds)$, in Section \ref{sec:EFT}, Eq. \eqref{eq:LW'}. 
Rotating to the fermion mass basis, the relevant $W'$ boson couplings are, up to small corrections due to EW symmetry breaking, given by
\begin{equation}
	c_q^{23} \approx \sin\theta_{b_R} \sin\theta_{c_R}~, \qquad c_N^{3} \approx \sin\theta_{\tau_R} \sin\theta_{N}.
\end{equation}
The corrections to $R(\dds)$ are maximised in the limit $c_q^{23}, c_N^{3} \to 1$, in which case Eq.~\eqref{eq:fit} implies $v_V \approx 1.8$~TeV in the minimal model, where all the breaking of $SU(2)_V\times U(1)'\to U(1)_Y$ is due to $H_V$.

\subsection{Neutrino masses} 
\label{sec:neutrinomasses:first}
The neutrino mass matrix, for a simplified case of a single SM-like neutrino flavor, has the following form in the basis $(\nu'_L, \nu'^{\,c}_R,N'^{}_L,N'^{c}_R)$,
\beq
{\cal M}_\nu=\left(\begin{array}{cccc}
0 & \frac{y_{\nu} v_{\rm EW}}{\sqrt{2}} & 0 & 0\\
\frac{y_{\nu} v_{\rm EW}}{\sqrt{2}} & \mu & \frac{\lambda_{\nu}v_{V}}{\sqrt{2}} & 0\\
0 & \frac{\lambda_{\nu}v_{V}}{\sqrt{2}} & 0 & M_{L}\\
0 & 0 & M_{L} & 0
\end{array}\right)~,
\label{eq:neutrinomass}
\eeq
where we have included a Majorana mass term $\mu$ for $\nu_R'$, which is a singlet under ${\cal G}$. For $v_{\rm EW}=0$, the SM neutrino $\nu_L'$ decouples from the system and remains massless.
In the remaining system of three Weyl fermions, the $\mu=0$ limit produces a massless Majorana neutrino $N_R^c=\cos\theta_{N} \nu_R'^{c}-\sin\theta_{N} N_R'^{c}$, where $\tan\theta_{N} = (\lambda_\nu v_V) / (\sqrt{2}M_{L})$, while the other two Weyl fermions combine into a Dirac fermion with mass 
\beq
\label{eq:MN':def}
M_{N'}\equiv M_L\sqrt{1 + \tan^2\theta_N}\,. 
\eeq
As with the charged fermions (discussed above), for $\lambda_\nu v_V\gg M_{L}$ the massless right-handed neutrino has a large admixture of $N_R'^c$, which is charged under $SU(2)_V$; this large mixing is necessary to induce a large coupling of the massless state to $W'$ in order to explain the $\mathcal{R}(\dds)$ anomaly. 
Introducing a nonzero but small $\mu \ll M_L,  {\lambda_{\nu} v_{V}}$ results in the lightest right-handed neutrino $N_R$ obtaining a mass $M_{N_R}\approx \mu \, (M_L / M_{N'})^2$ and a small admixture of $N_L'$. 
The heavy Dirac fermion becomes a pseudo-Dirac state, composed of two ${\mathcal O}(M_{N'})$ mass states split by ${\mathcal O}(\mu)$.

The above features persist for $y_\nu v_{\rm EW}\ne0$, i.e., when the SM $\nu_L'$ state is coupled to this system, in the phenomenologically interesting limit ${y_{\nu} v_{EW}} \ll \mu$. This also leads to a Type-I seesaw step that generates light Majorana neutrino masses $\approx y_\nu^2 v_{\rm EW}^2 / (2 \mu)$. It is straightforward to extend the above discussion to three generations of neutrinos, thereby accounting for the observed neutrino oscillation phenomena. In addition to the tree level neutrino masses discussed here, a Dirac mass term analogous to $y_\nu v_{\rm EW}$ is also generated at two loops. The size of this contribution depends on the flavor structure of the theory, which will be discussed in the next section. Hence we postpone a discussion of the two loop Dirac mass term, along with the discussion of the phenomenology of the additional neutrino states, until Section~\ref{sec:neutrino}.

%%%%%%%%%%%%%%%%%%%%%%%%%%%%%%%%%%%%%%%%%%%%
\section{Constraints}
\label{sec:constr}

In this section we derive the phenomenological constraints on the `3221' model. In addition to the SM states, 
the minimal model contains a light right-handed neutrino, $N_R$, and several heavy states: the vectorlike quarks, $U$ and $D$, with charges $2/3$ and $-1/3$, the charged lepton, $E$, and a heavy pseudo-Dirac neutrino, $N_H$, a heavy Higgs scalar $h_V$ from the $SU(2)_V$ Higgs doublet, $H_V$; and the $W'$ and $Z'$ gauge bosons. We also extend this minimal set-up by including up to two additional copies of vector-like fermions, requiring that mixings of the additional vector-like quarks with the SM model fermions are negligible (but large enough that they decay promptly and do not lead to displaced vertices). 
In Appendix \ref{sec:beyondminimal} we then also discuss the implications of non-minimal $SU(2)_V$ breaking sectors.

\subsection{LHC constraints}
In general, we expect the most important LHC constraints to arise from the resonant production of $W'$ and $Z'$ gauge bosons, and from the pair production of the heavy vectorlike quarks, $U$ and $D$. The cross sections for $pp\to W', Z'$ production depend crucially  on the assumed flavor structure of the couplings. The couplings of $W'$ to SM quarks are induced from mixing of the light right-handed fermions, $d_R'^i, u_R'^i $ with $Q_L'$, which is a doublet of $SU(2)_V$. The couplings of SM quarks to $Z'$ arise from  $d_R'^i, u_R'^i $ and $Q_L'$ gauge quantum numbers under $SU(2)_V\times U(1)'$. 
 Similarly, $e_R'^i$ and $\nu_R'^i$ mix with $L_L'$, see Eq. \eqref{eq:Yukawas}.
The resulting interaction Lagrangian is
\begin{equation}
\label{eqn:intL}
\begin{split}
	\mathcal{L}& \supset \frac{g_Y}{s_V c_V} \Big(\tilde c_d^{ij} \bar d_R^i\slashed Z' \negmedspace d_R^j + \tilde c_u^{ij} \bar u_R^i\slashed Z' \negmedspace u_R^j \Big) +\Big( \frac{g_V}{\sqrt{2}}  c_q^{ij} \bar u_R^i\slashed W' \negmedspace d_R^j
+{\rm h.c.}\Big)	\\
	 &+\frac{g_Y}{s_V c_V}\Big(\tilde c_e^{ij}  \bar e_R^i \slashed Z' \negmedspace e_R^j  +   \tilde c_N^3  \bar N_R \slashed Z' \negmedspace N_R\Big)  + \Big(\frac{g_V}{\sqrt{2}} c_N^{i}  \bar N_R^i\slashed W' \negmedspace e_R +{\rm h.c.}\Big).
\end{split}	 
\end{equation}
For couplings to right-handed charged leptons we take for the Yukawa in Eq. \eqref{eq:Yukawas}
\beq
\label{eq:lambdae}
\lambda_e^i\sim (0,0,1),
\eeq
so that there are no FCNCs induced among SM leptons at tree level, and the mixing is only among $E_R$ and $\tau_R$. We take $v_V \gg M_L$, so that the heavy mass eigenstate has a mass  ${\mathcal O}( v_V)$. The light eigenstate has a mass $m_\tau=y_\tau' v_{\rm EW}/\sqrt{2}$, where $y_\tau'= y_\tau/\sqrt{1+v_V^2/(2M_L^2)}$, is the SM $\tau$ Yukawa, with $y_\tau$ the coupling in  \eqref{eq:SMYukawa}. The mixing angle between right-handed $\tau$ and $E_R$ is $\sin\theta_{\tau_R}\sim {\mathcal O}(1)$, while the mixing among the left-handed $\tau$ and $E_L$ is highly suppressed, $\sin\theta_{\tau_L}\sim {\mathcal O}(m_\tau/v_V)$. One can allow for ${\mathcal O}(1)$ factors in  eq.!\eqref{eq:lambdae} which we absorb in the definition of $c_N^{i}$ and write in the numerical analysis 
\beq
c_N^{i}={\rm diag}(0,0,1).
\eeq
Note that the couplings of $Z'$ and $W'$ that involve the SM neutrinos are small and can be ignored. 

If one were able to expand in $v_V/M_Q$, the couplings in \eqref{eqn:intL} would be
\beq\label{eq:tildec}
\tilde c_d^{ij}\propto \frac{v_V^2}{M_Q^2}\lambda_d^i\lambda_d^j, \qquad 
\tilde c_u^{ij}\propto \frac{v_V^2}{M_Q^2} \lambda_u^i\lambda_u^j, \qquad 
c_q^{ij}\propto\frac{v_V^2}{M_Q^2}\lambda_u^i\lambda_d^j\,.
\eeq
 This illustrates how the hierarchy in $\lambda_{d,u}^i$ translates to a hierarchical structure of the couplings of SM quarks to $Z'$ and $W'$ gauge bosons. In the numerical analysis we work in a different limit, $v_V \gg M_Q\sim {\mathcal O}(v_{\rm EW})$.  This introduces a new dimensionless ratio $M_Q/v_V$, that needs to be taken into account. We show first  the results for the   minimal set of nonzero Yukawa couplings, $\lambda_d^i, \lambda_u^i$, Eq. \eqref{eq:Yukawas}, in order to explain the $R(\dds)$ anomaly. We then modify this minimal assumption and show the relevant constraints from FCNCs.

We first fix the flavor structure to a particular realization of the flavor-locking mechanism, see App. \ref{app:FL}, giving us the ``flavor-locked $23$ model'' (FL-23). The new states only couple to $c_R$ and $b_R$ in the mass eigenstate basis, so that 
\beq
\label{eqn:FL23}
\lambda_d^{i}\sim (0,0,1),\qquad {\rm and~~} \lambda_u^i\sim (0,1,0).
\eeq
As stated before, we are interested in the limit $v_V\gg m_Q$. For concreteness, we take $m_Q/v_V\sim \lambda$, the usual Wolfenstein CKM parameter, and $m_Q\sim v_{\rm EW}$. In this case one obtains for the couplings in eq.~\eqref{eqn:intL}
\beq
\label{eq:FL23:scaling}
\begin{split}
\tilde c_d^{ij}&\sim \Big(\frac{1}{2}+s_V^2\frac{1}{6}\Big)
\begin{pmatrix}
0 & 0 &0\\
0 & 0 &0\\
0 & 0 & 1
\end{pmatrix}
+\frac{1}{3} s_V^2 
\begin{pmatrix}
1 & 0 &0\\
0 & 1 &0\\
0 & 0 & \lambda^2
\end{pmatrix},
\\
\tilde c_u^{ij}&\sim 
\Big(\frac{1}{2}-s_V^2\frac{1}{6}\Big)
\begin{pmatrix}
0 & 0 &0\\
0 & 1 &0\\
0 & 0 & 0
\end{pmatrix}
+\frac{2}{3} s_V^2 
\begin{pmatrix}
1 & 0 &0\\
0 & \lambda^2 &0\\
0 & 0 & 1
\end{pmatrix}, \qquad\text{[FL-23]},
\\
c_q^{ij}&\sim
\begin{pmatrix}
0 & 0 &0\\
0 & 0 &1\\
0 & 0 & 0
\end{pmatrix}.
\end{split}
\eeq
Note that in \eqref{eq:FL23:scaling} the $\bar b \slashed W' \negmedspace c$ coupling is $c_q^{23}\sim 1$, which is parametrically larger than the corresponding CKM matrix element in the SM, $V_{cb}\sim \lambda^2$. 
  Similarly, the $Z'$ couples most strongly to charm and bottom quarks, with $\tilde c_u^{22}\sim {\mathcal O}(1)$ and $\tilde c_d^{33}\sim {\mathcal O}(1)$.  
  
For the FL-23 flavor structure the most severe LHC constraints are due $W'$ productions, decaying through $W'\to \tau N_R$, and from the $Z'$ production, decaying through the $Z' \to \tau\tau$.
Following the FL-23 setup we assume
in the numerical analysis of the LHC constraints that: {\it i)} there are sizeable mixings of vector-like fermions with $b'_R$, $c'_R$, $s'_R$ and $\nu'_R$, i.e., that $\sin\theta_{b_R, c_R, \tau_R, N} \simeq 1$,
 and {\it ii)} the  mixings with the other SM fermions are negligible, as in eq.~\eqref{eq:FL23:scaling}. The LHC constraints from $pp\to Z'\to \tau \tau,$ and $pp\to W'\to \tau N_R$ also depend crucially on 
 how many other channels besides the one containing $\tau$ leptons are open. If only the decay channels to SM quarks are open, the $W'$ branching ratios in the FL-23 model are 
 $Br(W'\to \tau N_R):Br(W'\to c b)\simeq 1:3$. In this case the LHC bounds from $pp\to W'\to \tau+$MET are severe enough, that the model is pushed close to the perturbative limit. 
 The situation changes, however, if vector-like fermions are light enough that $Z'$ and $W'$ can decay into them. 

\begin{figure}[t]
\begin{center}
\includegraphics[width=7.5cm]{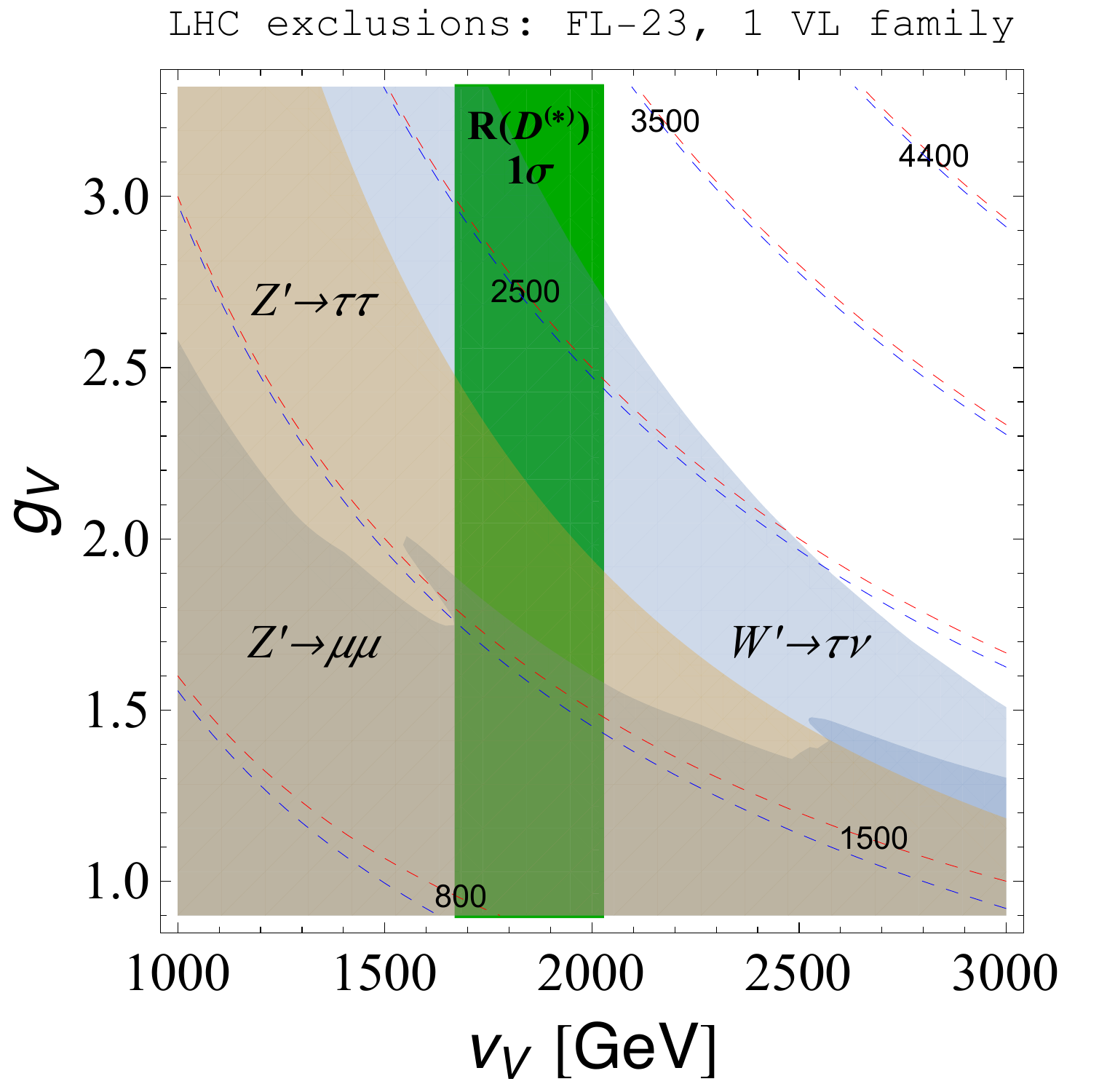} \includegraphics[width=7.5cm]{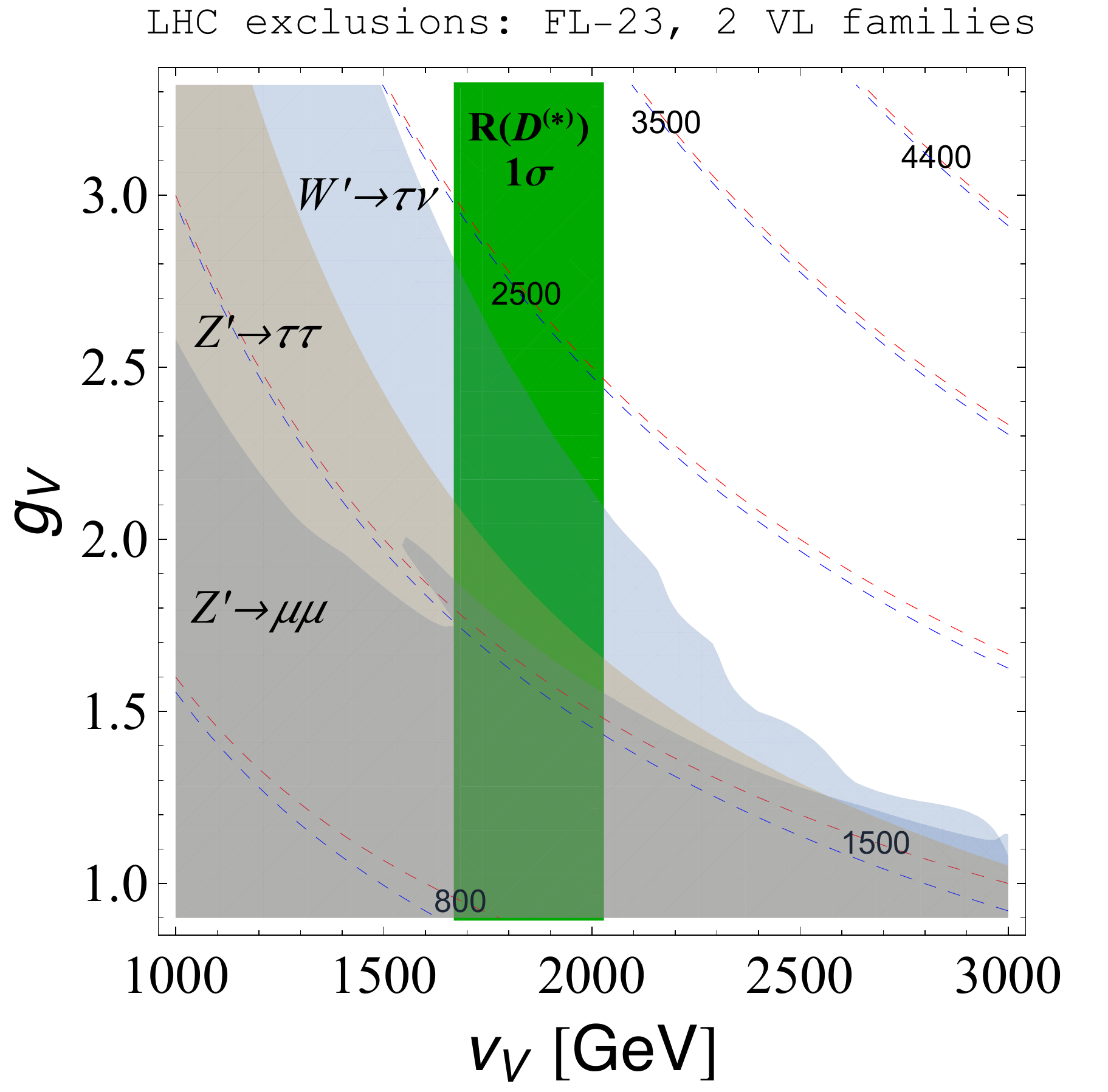}
\end{center}
\caption{The LHC exclusion limits on the $Z'$ and $W'$ resonances from ATLAS  $\tau^+\tau^-$~\cite{Aaboud:2017sjh}, $\ell^+ \ell^-$~\cite{Aaboud:2017buh} ($\ell=e,\mu$), and $\tau \nu$~\cite{Aaboud:2018vgh} searches, respectively, projected on the $(v_V, g_V)$ plane for the FL-23 scenario assuming the maximal fermion mixing angles $s_{\theta_b}$, $s_{\theta_c}$, $s_{\theta_e}$ and $s_{\theta_N}$ (that is $c_q^{23}, c_N^{3} \to 1$). The vertical green band represents $1\sigma$ range for $R(\dds)$ anomaly. Dashed blue (red) isolines are the predicted masses for $Z'$ ($W'$) gauge bosons. The plot on the left is for the minimal matter content, while the plot on the right assumes an additional family of the vector-like fermions mixing weakly with the SM fermions. Their masses are set to $0.8$~TeV, above the limits from~\cite{Sirunyan:2017lzl}. }
\label{fig:LHCZp}
\end{figure}

In Fig. \ref{fig:LHCZp} we show two examples, for one (left plot), and two (right plot) pairs of vector-like fermions. Comparing the  $\sigma(p p \to Z') \times \mathcal{B}(Z'\to \bar f f)$ the upper limits from the ATLAS  $\tau^+\tau^-$~\cite{Aaboud:2017sjh} and $\ell^+ \ell^-$~\cite{Aaboud:2017buh} ($\ell=e,\mu$) searches gives the exclusion regions in the $(v_V, g_V)$ plane shown in Fig.~\ref{fig:LHCZp} for $\tau^+\tau^-$ (brown) and $\ell^+ \ell^-$ (gray), respectively. 
The parameter space consistent with the LHC data has $g_V \gg g'$, or $t_V \ll 1$. This is required to suppress $Z'$ couplings to valence quarks and light charged leptons. In this regime, the dominant decay modes are to $b \bar b$, $c \bar c$, $\tau^+ \tau^-$ and $N_R N_R$, and the main production mechanism is from the charm fusion. 
Comparing instead the $\sigma(p p \to W') \times \mathcal{B}(W'\to \tau \nu)$ to the upper limits from the ATLAS analysis~\cite{Aaboud:2018vgh} (see also~\cite{CMS:2016ppa}), leads to constraints shown with light blue.
Introducing another vector-like fermion family helps reduce these constraints as shown in the right plot.  Here we set the masses of vector-like fermion to $0.8$~TeV, which is above the limits from the quark partner pair production~\cite{Sirunyan:2017lzl}. We also checked that in in the interesting region of parameter space the $W', Z'$ induced production is always subleading compared to the QCD pair production.

\subsection{Flavor constraints}
We next turn our attention to the flavor constraints. In FL-23 model all the tree-level FCNCs are strongly suppressed, and are phenomenologically negligible. The one-loop induced FCNCs are also negligible, suppressed by both  $m_{W'}\gg m_W$ and the extreme smallness of the flavor-changing couplings $c_q^{ij}$, for $ij\ne 23$. 
  
Other flavor models, beside flavor-locking, may lead to a flavor structure similar to eq.~\eqref{eq:FL23:scaling}, but with vanishing entries modified to some nonzero value.  
In fact, when writing eq.~\eqref{eq:FL23:scaling} we assumed that the two SM Yukawa structures are aligned for the right-handed fields with the FL-23 spurions, i.e., that no right-handed rotations are needed to diagonalize them. If we assume instead that the SM flavor structure comes from a Froggatt-Nielsen (FN) flavor model with a single horizontal $U(1)$ \cite{Leurer:1992wg,Leurer:1993gy}, 
while the couplings to vector-like fermions are due to FL-23,  some of the vanishing entries become nonzero. The largest correction to the vanishing entries in this case is in $\tilde c_{u}^{23}\sim {\mathcal O}(\lambda^{4})$, with $\lambda=0.23$ the CKM parameter, and is ${\mathcal O}(\lambda^8)$ or less in all the other cases, all of which can still be safely ignored.

The off-diagonal $Z'$ couplings induce tree-level FCNCs which are stringently constrained by the bounds on the $B\to K^{(*)}\nu\nu$ branching ratios and by the measurements of $B_{d,s}-\bar B_{d,s}$, $D^0-\bar D^0$ and $K^0-\bar K^0$ mixing amplitudes. The constraints from $B_c\to \tau \nu$ were already discussed in eq.~\eqref{eq:Bctaunu}, and were shown to be satisfied in these types of models. 

The branching ratio for $B\to K^{(*)} \bar N_R N_R$ normalized to the SM value for $B\to K^{(*)} \bar \nu_\ell \nu_\ell$, is given by
  \beq
  \begin{split}
 R_{K^{(*)}\nu\nu}\equiv \frac{Br(B\to K^{(*)} \nu\nu)}{Br(B\to K^{(*)} \bar \nu_\ell \nu_\ell)|_{\rm SM}}=&
 1+\frac{1}{6}\left(
 \frac{g_Y^2}{s_V^2 c_V^2} \frac{\tilde c_D^{23} \tilde c_N}{m_{Z'}^2}
  \frac{\pi s_W^2}{G_F \alpha \big|V_{tb} V_{ts}^*\big| X(x_t)}
  \right)^2
  \\
  &\simeq 1+1.9 \left(\frac{0.3}{s_V}\right)^4\left(\frac{2{\rm~TeV}}{m_{Z'}}\right)^4\left(\frac{\tilde c_d^{23} \tilde c_N^3}{\lambda^2}\right)^2,
  \end{split}
  \eeq
  where $G_F=1.166 378 7(6)\times 10^{-5} {\rm GeV}^{-2}$ is the Fermi constant, $|V_{tb}|\simeq 1,  |V_{ts}|=40.0(2.7)\times 10^{-3}$ are the CKM elements, $s_W^2\simeq 0.231$ is the square of the sine of the weak mixing angle, $\alpha=1/137$ the fine-structure constant, and $X(x_t)\simeq 1.31$ the loop function. The present experimental bound is $R_{K^{(*)}\nu\nu} <5.2$ at 95\% C.L. \cite{Patrignani:2016xqp}, which signifies that for $m_{Z'}\sim 2$ TeV one requires $\tilde c_{d}^{23} \tilde c_N^3\lesssim \lambda^2$. A suppression of this size is usually not a challenge for flavor models that have suppressed FCNCs.
  
  The branching ratio for $D_s\to \tau N_R$ normalized to the SM prediction for $D_s\to \tau \nu_\tau$ is given by
  \beq
  \begin{split}
  R_{D_s\tau\nu}=&\frac{Br(D_s\to \tau \nu)}{Br(D_s\to \tau \nu_\tau)|_{\rm SM}}=\left(\frac{g_V^2}{g^2}\frac{c_q^{22} c_N^3}{V_{cs}}\frac{m_W^2}{m_{W'}^2}\right)^2
  \\
  =&1+3.2 \times 10^{-4} \big(g_V^2 c_q^{22} c_N^3)^2 \left(\frac{1{\rm TeV}}{m_{W'}}\right)^4.
  \end{split}
  \eeq
The correction is well below the present experimental precision on this branching ratio, $Br(D_s\to \tau \nu)=5.48(23)\times 10^{-2}$, even for $c_q^{22} c_N^3\sim {\mathcal O}(1)$.

The most severe bounds arise from the absence of any deviations seen in the meson mixing measurements.  The contributions to the meson mixing from tree-level $Z'$ exchanges can be parametrized by the effective Hamiltonian 
\beq
H_{\rm eff}=\tilde C_1^{q_i q_j}  \tilde Q_1^{q_i q_j}, 
\eeq
where $\tilde Q_1^{q_i q_j}=(\bar q_i \gamma_\mu P_R q_j)^2$ \cite{Bona:2007vi}. The bounds on the Wilson coefficient $\tilde C_1^{ij}$, are \cite{Bevan:2014cya,Martinelli:2015}
\begin{subequations}
\begin{align}
1/|\tilde C_1^{sd}| &> \big(1 \times 10^{3} {\rm ~TeV}\big)^2,   \qquad{\rm Re}(C_K),
\\
1/|\tilde C_1^{sd}| & > \big(2 \times 10^{4} {\rm ~TeV}\big)^2,  \qquad {\rm Im}(C_K),
\\
1/|\tilde C_1^{cu}| &>  \big(6 \times 10^{3} {\rm ~TeV}\big)^2, \qquad{\rm Im}(C_D),
\\
1/|\tilde C_1^{bd}| &> \big(9\times 10^{2} {\rm ~TeV}\big)^2,  \qquad ~C_{B_d},
\\
1/|\tilde C_1^{bs}| &> \big(2\times 10^{2} {\rm ~TeV}\big)^2, \qquad ~C_{B_s},
\end{align}
\end{subequations}
where the bounds are due to the allowed size of the real and imaginary parts of the NP Wilson in $\epsilon_K$, the bound on weak phase in $D-\bar D$ mixing and on the size of NP matrix elements in $B_{d(s)}-\bar B_{d(s)}$  mixing, in all cases factoring out the SM weak phase.

The tree level $Z'$ exchange gives for the Wilson coefficients of the NP operators
\beq
\tilde C_1^{q_i q_j}=\left(\frac{g_Y \tilde c_q^{ij}}{s_V c_V}\right)^2 \frac{1}{m_{Z'}^2}.
\eeq
The meson mixing bounds translate to
\begin{subequations}
\begin{align}
&\big|\tilde c_d^{12}\big|\lesssim 0.9 \cdot \lambda^4 \cdot  \left(\frac{s_V}{0.3}\right) \cdot \left(\frac{ m_{Z'}}{3{\rm~TeV}}\right),
\qquad~~~{\rm Re}(C_K),
\\
&\big|\tilde c_d^{12}\big|\lesssim 0.8 \cdot \lambda^6 \cdot  \left(\frac{s_V}{0.3}\right) \cdot \left(\frac{ m_{Z'}}{3{\rm~TeV}}\right), \qquad ~~~{\rm Im}(C_K),
\\
&\big|\tilde c_u^{12}\big|\lesssim 0.6 \cdot \lambda^5 \cdot  \left(\frac{s_V}{0.3}\right) \cdot \left(\frac{ m_{Z'}}{3{\rm~TeV}}\right), \qquad~~~{\rm Im}(C_D),
\\
&\big|\tilde c_d^{13}\big|\lesssim 0.9 \cdot \lambda^4 \cdot  \left(\frac{s_V}{0.3}\right) \cdot \left(\frac{ m_{Z'}}{3{\rm~TeV}}\right), \qquad ~~~C_{B_d},
\\
&\big|\tilde c_d^{23}\big|\lesssim 1.0 \cdot \lambda^3 \cdot  \left(\frac{s_V}{0.3}\right) \cdot \left(\frac{ m_{Z'}}{3{\rm~TeV}}\right), \qquad ~~~C_{B_s},
\end{align}
\end{subequations}
where the CKM parameter $\lambda=0.23$. The required suppressions of $\tilde c_q^{ij}$ are highly non-trivial, and would, e.g., be violated in most realizations of, otherwise phenomenologically viable, FN models. 

Finally, the corrections to electroweak observables from heavy vectorlike fermions and due to $W-W'$ mixing are well below present experimental sensitivity. Since $v_V\gg v_{\rm EW}$ the vectorlike fermions are heavy ${\mathcal O}(1 {\rm~TeV}$), see Eq. \eqref{eq:MD}. The corrections to $T$ parameter from $W-W'$ mixing arise effectively at 2-loops and are further suppressed by the $m_{W'}$ mass.

%%%%%%%%%%%%%%%%%%%%%%%%%%%%%%%
\section{Neutrino Phenomenology}
\label{sec:neutrino}
%%%%%%%%%%%%%%%%%%%%%%%%%%%%%%%

In this section, we study the phenomenology associated with the sterile neutrinos that are part of our framework. To simplify the discussion we assume that, as for the charged states, only one pair of vector-like fermions mixes appreciably with the SM fermions through Yukawa interactions. In addition to the three generations of SM neutrinos the relevant fields are thus the vector-like fermion pair $N_L', N_R'$ and the three singlet right handed neutrinos $\nu_R^i$. These give rise to the following mass eigenstates:

\begin{itemize}
\item $N_R$ is an admixture of $\nu_R^i$ and $N_R'$ with mass $M_{N_R}\approx \mu \, (M_L / M_{N'})^2$, with $\mu, M_L, M_{N'}$ defined in Eqs. \eqref{eq:neutrinomass}, \eqref{eq:MN':def}. It couples appreciably to the $W'$ and is responsible for the $R(\dds)$ signal. For simplicity we take $i=1$, i.e., $\mu^1=\mu$, and treat the mass $M_{N_R}$ as a free parameter.
\item $\nu_R^{2,3}\approx \nu_R'^{ 2,3}$ are the remaining two singlets. We assume that they couple negligibly to $N_L, N_R$ and are approximately degenerate, so that they have masses $M_{\nu_R^{2,3}}\approx \mu^{2,3}\approx \mu$. These states are therefore expected to be heavier than $N_R$ by a factor $\sim (M_{N'}/M_L)^2$. We will use these states to generate the observed neutrino masses via  type-I seesaw mechanism, giving $m_{\nu_L}\approx y_\nu^2 v_{\rm EW}^2 / (2 \mu)$.\footnote{This requirement imposes requirements on the Yukawa couplings of $\nu_R'^{ 2,3}$ and mixing angles with SM neutrinos. Since solar and atmospheric neutrino oscillation data only fix two mass differences, while the absolute mass scale for active neutrinos is only bounded from above, only two sterile neutrinos are required to participate in the seesaw relation. The remaining sterile neutrinos, including $N_R$ can in principle be decoupled from the seesaw constraint.}
\item The remaining two degrees of freedom make up a pseudo-Dirac state composed of two states with masses of ${\mathcal O}(M_{N'})$ split by ${\mathcal O}(\mu)$. These are heavy and decay rapidly, hence do not directly influence the low energy neutrino phenomenology and the cosmology, and thus do not discuss them further. 
\end{itemize}

\begin{figure}[t]
\begin{center}
\includegraphics[width=7.5cm]{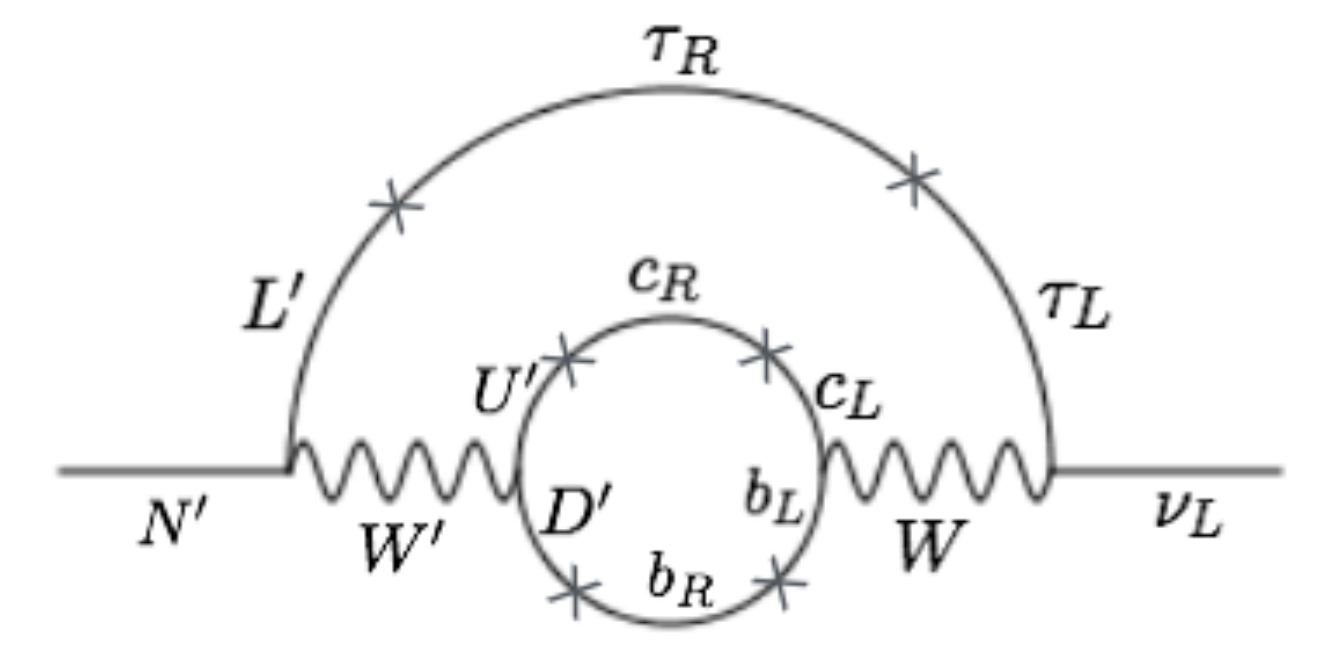}
\end{center}
\caption{The two loop contribution to Dirac mass for neutrinos. Crosses on fermion propagators denote mass insertions from $\vev{H_V}$ or $\vev{H}$, as inferred from the Lagrangian. }
\label{fig:twoloop}
\end{figure}

In our setup a Dirac mass term is generated at two loops, and is sensitive to the flavor structure of the theory, see Fig.\,\ref{fig:twoloop}. This contribution has been approximately estimated in \cite{Babu:1988yq,Balakrishna:1988bn,Borah:2017leo}. Ignoring $\mathcal{O}(1)$ pre-factors and integration functions, the Dirac mass is in the FL-23 scenario approximately given by
\beq
m_D\sim \frac{g^2 \, V_{cb}}{512 \pi^4}\,\frac{C_{23,3}}{\Lambda_{\text{eff}}^2}\,m_b\,m_c\,m_\tau \approx \mathcal{O}(10^{-3})\, \text{eV}.
\eeq
This is much smaller than the active neutrino mass scale, and therefore does not modify the discussions of neutrino masses and mixings above.

\subsection{Cosmology}

The same $W'$ mediated interaction that gives the $R(D^{*})$ signal will also produce $N_R$ in the early Universe, e.g., through the processes $bc\to \tau N_R$, or $\tau\tau\to N_R N_R$. These thermalize the $N_R$ population with the SM bath at high temperatures. Once the temperature drops below the masses of the SM fermions involved in these interactions, the $N_R$ abundance freezes out. Since we have assumed $m_{N_R}\lesssim\, {\mathcal O}(100$ MeV), $N_R$ freezes out at temperature above $m_{N_R}$, so that its abundance is not Boltzmann suppressed. It thus survives as an additional neutrino species in the early Universe.

\begin{figure}[t]
\begin{center}
\includegraphics[width=6cm]{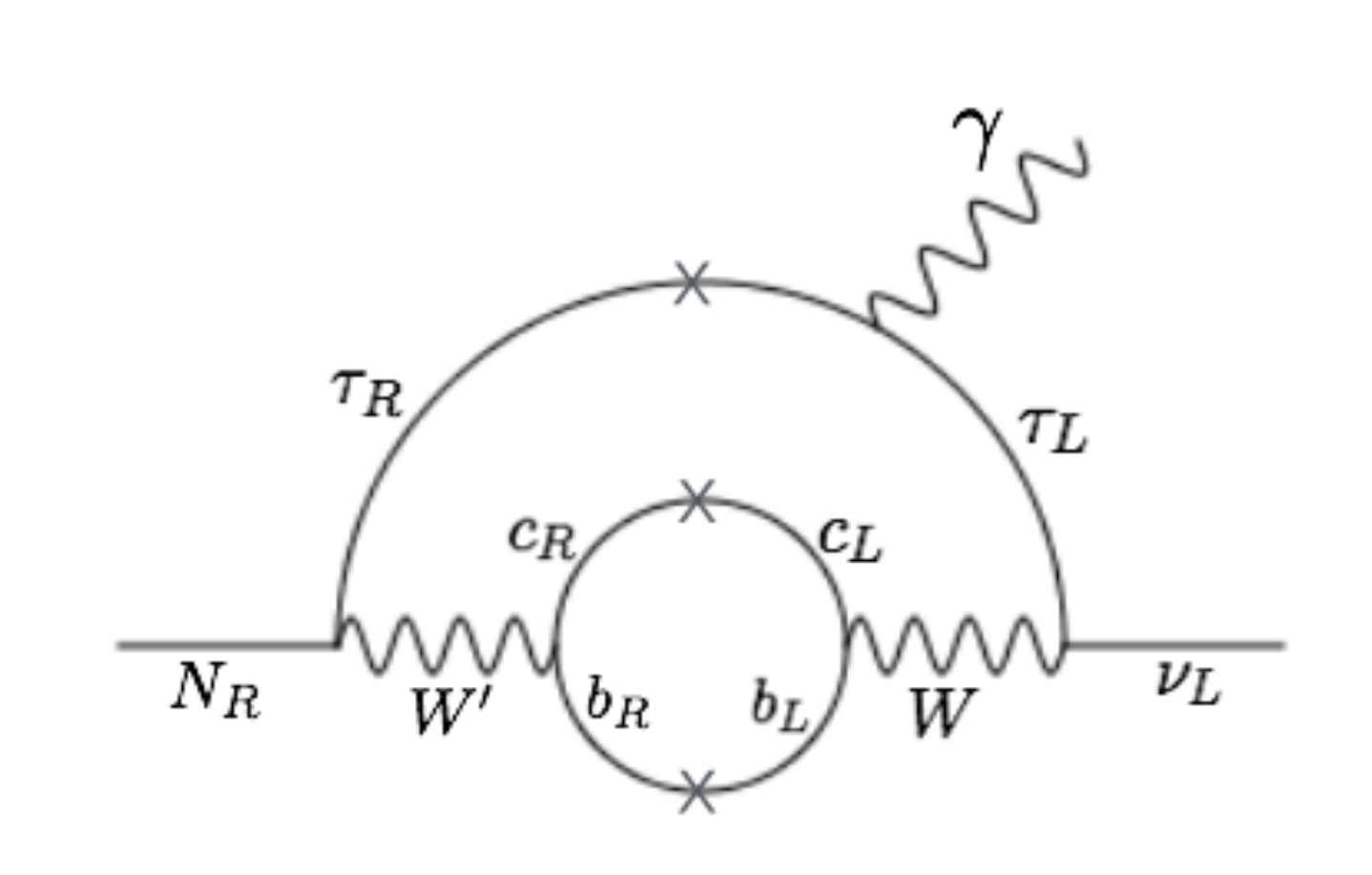}~~~~~~~~\includegraphics[width=5.5cm]{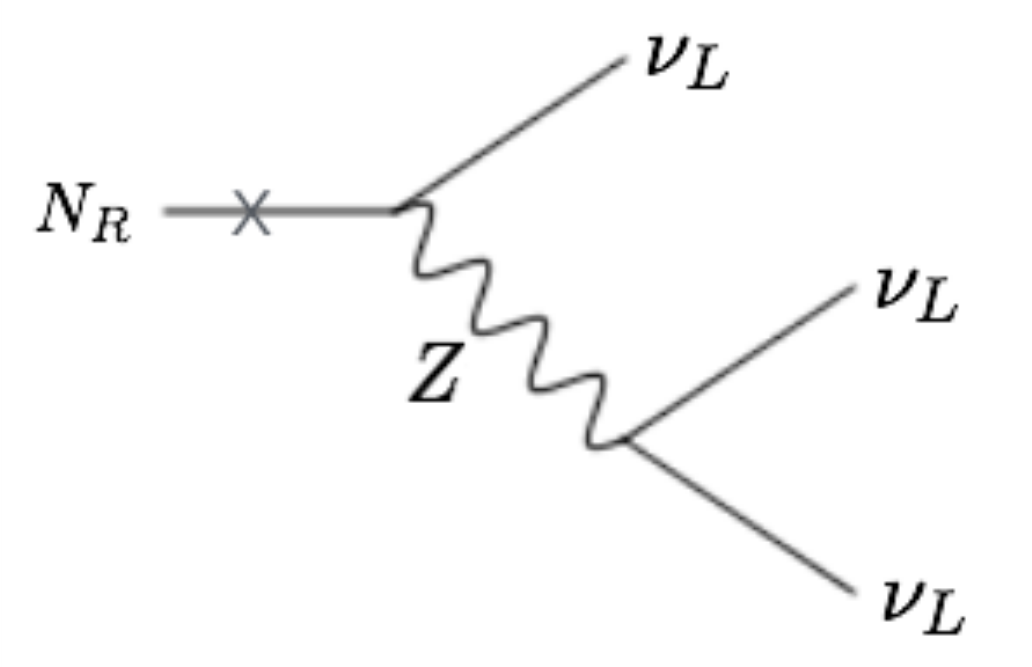}
\end{center}
\caption{Decay modes of $N_R$. The radiative decay (left) is induced by its coupling to $W'$ as dictated by the fit to the $R(\dds)$ signal. The tree level decay to neutrinos (right), induced by $N_R-\nu_L$ mixing, is the standard decay channel for sterile neutrinos.}
\label{fig:decays}
\end{figure}

It then becomes crucial to determine the fate of this $N_R$ population. The $N_R$ can decay either through $N_R\to \nu\gamma$ via a two loop radiative process induced by its $W'$ couplings, or via a small mixing with the SM neutrinos, see Fig.\,\ref{fig:decays}. Since the $N_R$ mixing angle with the SM neutrinos can be arbitrarily small, the radiative decay process is generally the dominant decay channel. The decay rate for this process is approximately \cite{Lavoura:2003xp,Wong:1992qa,Bezrukov:2009th}
\begin{align}
\Gamma_{N_R\to \nu\gamma}
	&\simeq\frac{\alpha}{32\, \pi^8}\, V_{cb}^2\, \frac{G_F^2\, m_\tau^2\, m_b^2\, m_c^2\, m_{N_R}^3}{(\Lambda_{\text{eff}}^2/C_{23,3})^2}\, \text{ln}\left(\frac{m_b^2}{m_c^2}\right)\nonumber\\
	&\simeq 10^{-49} \bigg(\frac{m_{N_R}}{\text{keV}}\bigg)^3 \text{GeV}.
\end{align}
It should be emphasized that this decay rate is $\emph{completely}$ fixed by the fit to $R(D^{(*)})$, as there are no other free parameters that enter the above decay rate. For comparison, the decay rate for the tree level process, Fig.\,\ref{fig:decays} right, is
\begin{align}
\Gamma_{N_R\to 3\nu}
	 & \simeq\frac{G_F^2}{192\, \pi^3}\, m_{N_R}^5 \sin^2\theta \nonumber\\
	&  \simeq 10^{-48} \bigg(\frac{m_{N_R}}{\text{keV}}\bigg)^5 \bigg(\frac{\sin^2\theta}{10^{-4}}\bigg)\,.
\end{align}
The mixing angle is bounded from above, sin$^2\theta\lesssim m_{\nu}/m_{N_R}$, in order to remain consistent with the seesaw mechanism, but is typically much smaller, rendering this mode subdominant.

The radiative decay channel $N_R\to \nu\gamma$, if dominant, corresponds to a lifetime of $\sim 10^{25} \left(m_{N_R}/{\text{keV}}\right)^{-3}$ s. For $m_{N_R}\,\textless\, \mathcal{O}(100)$ keV, the $N_R$ sterile neutrino therefore has a lifetime greater than the age of the Universe and could in principle form a component of dark matter. Such a dark matter interpretation, however, faces several challenges. 

It is well known that without other additional modifications of the standard cosmology, a species that undergoes relativistic freezeout overcloses the Universe, if its mass is greater than $\sim 100$ eV. Its relic abundance  can be made to match the observed dark matter abundance through appropriate entropy dilution. For instance, species that grow to dominate the energy density in the early Universe and decay late, after dark matter has frozen out, release significant entropy into the SM thermal bath and dilute the abundance of dark matter. Such long-lived particles are present in our framework in the form of $\nu_R^{2,3}$. If their masses lie at the GeV scale, they can thermalize, undergo relativistic freezeout, and decay just before BBN, diluting the abundance of dark matter by a factor of $\lesssim 30$ \cite{Scherrer:1984fd,Asaka:2006ek,Bezrukov:2009th}. Significantly larger dilution factors can be achieved with late decaying sterile neutrinos that are not part of the seesaw mechanism (see e.g. \cite{Bezrukov:2009th}), although these are not as well motivated in general. It should be noted that a large entropy dilution also helps to make the dark matter colder, making the light dark matter candidate more compatible with warm dark matter constraints.

Even with the correct relic abundance, dark matter in this mass range is severely constrained by $\gamma$-ray bounds from various observations \cite{Essig:2013goa}, which rule out dark matter lifetimes of $\mathcal{O} (10^{26-28})s$ in the keV-MeV window. These observations therefore rule out $N_R$,  which has a lifetime $\sim 10^{25} \left({m_{N_R}}/{\text{keV}}\right)^{-3}$\,s, as constituting all of dark matter. It could still constitute a small fraction, sub-percent level,  of dark matter, in which case future $\gamma$-ray observations could discover a line signal from its decay. This does requires significant entropy dilution of the dark matter abundance beyond what is possible in our framework. 

If $N_R$ is light, with a mass below $\lesssim$ keV, it can act as dark radiation and contribute to $N_{\text{eff}}$ at BBN and/or CMB decoupling. This is potentially problematic since a light sterile neutrino that undergoes relativistic freezeout and is long-lived effectively acts as an additional neutrino species, contributing $\Delta N_{\text{eff}}\approx1$, which is inconsistent with current observations. However, $\mathcal{O}(1)$ dilution of its abundance, as would be expected from $\nu_R^{2,3}$ decays, if they are at the GeV scale, would result in $\Delta N_{\text{eff}}\approx\mathcal{O}(0.1)$, which would be consistent with current observations and at the same time possibly within reach of future measurements. 

Alternatively, when $N_R$ is heavy enough that its lifetime is shorter than the age of the Universe, $N_R\to \nu\gamma$ as the dominant decay channel results in a late injection of photons into the Universe, which can distort the CMB or contribute to the diffuse photon background. This problem can be avoided by enhancing the $N_R$ mixing with active neutrinos, to the extent allowed by the seesaw mechanism, so that $N_R$ primarily decays via this mixing (into channels such as $N_R\to 3\nu$, see Fig.\,\ref{fig:decays} right). For $m_{N_R}\textgreater$ MeV, this introduces dominant decays channels into electrons or pions, which can also distort the CMB or contribute to the diffuse photon background. For masses below an MeV, $N_R\to 3\nu$ is the only available channel, which might be compatible with all existing constraints.

\subsection{Direct Production of Additional Sterile Neutrinos}

The above discussion suggests that the sterile neutrinos $\nu_R^{2,3}$ might be light, at the GeV scale, such that their late decays dilute the abundance of  $N_R$ in order to evade various cosmological constraints. This gives rise to the fascinating possibility that $\nu_R^{2,3}$ can be directly produced. Since they decay with lifetimes $\lesssim 1$ s, their decays can lead to observable direct signatures. Note that production of $\nu_R^{2,3}$ requires them to carry small admixtures of $\nu'_L$, which couples them to electroweak gauge bosons, or of $N'_L, N'_R$, which couples them to $W', Z'$ gauge bosons, as $\nu_R'^{2,3}$ are singlets under $\mathcal{G}$.

If they carry small admixtures of $N'_L, N'_R$, the $\nu_R^{2,3}$ states can be produced in place of $N_R$ in $B$ decays if kinematically allowed. The branching ratio is suppressed by the mixing angle with $N'_L, N'_R$ as well as by the phase space. Although these states still appear as missing energy, as for the $b\to c \tau N_R$ decay, the distribution of visible final states will be affected by the relatively heavy masses of $\nu_R^{2,3}$. Finally, $\nu_R^{2,3}$ can also be produced from the decays of $W'$ and $Z'$ at the LHC. Their relative long lifetimes $\lesssim 1$ s could then lead to displaced decay signals  at the LHC as well as at proposed detectors such as SHiP~\cite{Anelli:2015pba}, MATHUSLA~\cite{Chou:2016lxi}, FASER~\cite{Feng:2017uoz} or CODEX-b~\cite{Gligorov:2017nwh}.

\section{Conclusions}
\label{sec:conclusions}
In the present manuscript we discussed the possibility that the $R(D^{(*)})$ anomaly is due to an additional right-handed neutrino, giving rise to the $b\to c\tau N_R$ decay through an exchange of $W'$ that couples to right-handed currents. Since such a decay does not interfere with the SM $b\to c\tau \nu_\tau$ transition, it coherently adds to the $B\to D^{(*)}\tau \nu$ branching ratios, in agreement with the observed experimental trend. Assuming $N_R$ to have mass below $\mathcal{O}(100)$\,MeV, this additional channel leads to only negligibly small deviations in the kinematic distributions of the $B\to D^{(*)}\tau \nu$ decays.

The right-handed nature of the $W'$ interaction allows construction of a UV complete renormalizable model, based on extending the SM gauge group to $SU(3)_c\times SU(2)_L\times SU(2)_V\times U(1)'$. The flavor and collider searches constrain the model to have a definite flavor structure -- achievable in a flavor-locked framework -- and to also contain additional copies of vector-like fermions, to which the  $W'$ and $Z'$ bosons can decay. In this way the model becomes very predictive. The additional vector-like fermions cannot be too heavy. For $W'$ and $Z'$ with the mass of about 3 TeV the model becomes non-perturbative, since the two resonances become very wide. A clear prediction is therefore that there should be vector-like fermions with a mass below about 1.5 TeV. 

Another set of predictions is related to neutrino phenomenology. The sterile neutrino $N_R$ is light and long-lived, and has significant relic abundance. Hence it can contribute measurably to $N_{\text{eff}}$ at both BBN and CMB, while its decay $N_R\to\nu\gamma$ could also constitute an observable signal for current and future experiments.  Likewise, the model also contains heavier sterile neutrinos, potentially in the GeV mass range; these could lead to additional signals, either in $B$ decays or in searches for displaced vertices. 

Irrespective of what the future of $R(D^{(*)})$ anomaly will be, we encourage the experimental collaborations to explore possible distortions of the kinematical distributions in semileptonic $B$ meson decays due to the heavy right-handed neutrino in the final state --- an option which goes beyond the short-distance new physics effects typically considered.

\acknowledgments JZ wishes to thank Stefania Gori for a smooth drive from Chicago to Cincinnati, and SAS for a viable wi-fi connection, both crucial for writing and completion of the paper. BS is likewise grateful to Wolfgang Altmannshofer and Malte Buschmann for smooth and productive drives to/from Chicago at crucial stages of this project. DR thanks Florian Bernlochner, Stephan Duell, Zoltan Ligeti and Michele Papucci for their ongoing collaboration in the development of \texttt{Hammer}, which was used for part of the analysis in this work. DR and BS acknowledge support from the University of Cincinnati. JZ acknowledges support in part by the DOE grant DE-SC0011784. This work was performed in part at the Aspen Center for
Physics, which is supported by National Science Foundation
grant PHY-1066293.
 
\appendix

\section{Flavor-locked couplings}
\label{app:FL}
Although the flavor structure of the Yukawa couplings in Eq.~\eqref{eqn:FL23} can be treated as an ansatz, they may also arise dynamically in a flavor-locking (FL) context \cite{Knapen:2015hia,Altmannshofer:2017uvs}, hence the terminology used in the main text.  In the general setup of the FL mechanism for the SM, one posits the existence of three up and down type flavons $y_{\al = u,c,t} \sim \bm{3}\times\bar{\bm{3}}\times\bm{1}$ and $y_{\hal = d,s,b}  \sim \bm{3}\times\bm{1}\times\bar{\bm{3}}$, with respect to the flavor symmetry $U(3)_Q \otimes U(3)_U \times U(3)_D$. Each flavon carries typically also a unique $U(1)_{\alpha, \hat\alpha}$ (or a discrete symmetry), which is broken by `hierarchon' operators, that gain vevs. The vacuum of the flavon potential ensures that the up and down type flavon vevs $\langle y_{\al,\hal} \rangle$ are aligned, rank-1 and disjoint. That is, one has a 3-way portal
\begin{equation}
	\bar{Q}_L^i \frac{{y_\al}_{iJ}}{\Lambda_{\text{F}}} \frac{s_\al}{\Lambda_{\text{H}}} \tilde H U_R^J + \bar{Q}_L^i \frac{{y_\hal}_{i\hat J}}{\Lambda_{\text{F}}} \frac{s_\hal}{\Lambda_{\text{H}}} H D_R^{\hat J}\,,
\end{equation}
in which dynamically $\langle y_{u} \rangle = \text{diag}\{r_y,0,0\}$,  $\langle y_{c} \rangle = \text{diag}\{0,r_y,0\}$, $\langle y_{t} \rangle = \text{diag}\{0,0,r_y\}$, and similarly for $\langle y_{\hal = d,s,b} \rangle$, and the fermion mass hierarchies are controlled by $\langle s_{\al,\hal}\rangle$. The CKM is a flat direction of this potential, but may be lifted to a realistic flavor structure by the introduction of additional physics in the Higgs sector~\cite{Altmannshofer:2017uvs}. Here we assume that the dynamics of the $y_{\alpha, \hat\alpha}$ flavons is fixed to SM structure at a relatively high scale, and explore the dynamical generation of associated flavor violating couplings involving the $W'$ and $Z'$.

The case of interest for the $b\to c\tau\nu$ anomaly is  when there are two
additional flavons, $\lambda_u \sim \bm{1}\times\bar{\bm{3}}\times\bm{1}$ and $\lambda_d \sim \bm{1}\times\bm{1}\times\bar{\bm{3}}$, which can then appear in the $W'$ and $Z'$ couplings to SM quarks, that is in the operators
\begin{equation}
	\label{eqn:QZWpI}
	\mathcal{O}_{W',Z'} \sim \frac{\lambda^{i*}_u \lambda^j_u}{\Lambda^2} \bar{u}^i_R \slashed{Z}' u_R^j +  \frac{\lambda^{\hat\i*}_d \lambda^{\hat\j}_d}{\Lambda^2} \bar{d}^{\hat\i}_R \slashed{Z}' d_R^{\hat\j} + \frac{\lambda^{i*}_u \lambda^{\hat\j}_d}{\Lambda^2} \bar{u}^i_R \slashed{Z}' d_R^{\hat\j} + \text{h.c.}\,,
\end{equation}
where $\Lambda$ is the scale connected with the dynamics of $\lambda_{u,d}$ flavons. The renormalizable potential for $\lambda_{u,d}$ has the general form
\begin{align}
	V & = \mu_u\big( \text{Tr}[\lambda_u^\pda \lambda_u^\da] - r_u^2)^2 + \mu_d\big( \text{Tr}[\lambda_d^\pda \lambda_d^\da] - r_d^2)^2 \nn \\
	& + \nu_1 \Big|\Tr[ \lambda_u^\pda \lambda_u^\da] + \Tr[ \lambda_d^\pda \lambda_d^\da] + \Tr[ y_\al^\da y_\al^\pda ] +  \Tr[y_\hal^\da y_\hal^\pda]  - r_u^2 - r_d^2 - r_y^2 - r_y^2 \Big|^2 \nn \\
	& + \nu_{\al} \Tr[\lambda_u^\da \lambda_u^\pda y_\al^\da y_\al^\pda] + \nu_{\hal} \Tr[\lambda_d^\da \lambda_d^\pda y_\hal^\da y_\hal^\pda]\,, \label{eqn:xiV}
\end{align}
noting that $\lambda_u^\da \lambda_u^\pda \sim \bm{3}\times\bar{\bm{3}}$ of $U(3)_U$ and similarly for $\lambda_d$. All the constants, $\mu_{u,d}$, $\nu_1$ and $\nu_{\alpha, \hat \alpha}$, are real.

The $\mu_{u,d}$ and $\nu_1$ terms enforce $|\langle \lambda_u \rangle| = r_u$ and $|\langle \lambda_d \rangle| = r_d$. Defining the diagonal matrix $D_1 = \text{diag}\{1,0,0\}$, then in the quark mass basis the operator~\eqref{eqn:QZWpI} can be rewritten in matrix form, without loss of generality,
\begin{equation}
	\label{eqn:QZWpI2}
	\mathcal{O}_{W',Z'} \sim  \frac{r_u^2}{\Lambda^2} \bar{u}_R \tilde{U}^\pda_u D_1 \tilde{U}^\da_u \slashed{Z}' u_R + \frac{r_d^2}{\Lambda^2} \bar{d}_R \tilde{U}^\pda_d D_1 \tilde{U}^\da_d \slashed{Z}' d_R + \frac{r_ur_d}{\Lambda^2}\bar{u}_R \tilde{U}^\pda_u D_1 \tilde{U}^\da_d \slashed{W}' d_R + \text{h.c.}\,,
\end{equation}
in which $\tilde{U}_{u,d}$ are unitary matrices, while simultaneously the $\nu_{\al,\hal}$ terms of the potential become
\begin{equation}
	\nu_{\al} r_u^2 r_y^2 \big|[\tilde{U}_{u}]_{\al 1}\big|^2 + \nu_{\hal} r_d^2 r_y^2 \big|[\tilde{U}_{d}]_{\hal 1}\big|^2\,.
\end{equation}
For $\nu_{\al,\hal} > 0$ and provided $\nu_{c} < \nu_{u,t}$ and $\nu_{b} < \nu_{d,s}$, unitarity of $\tilde{U}_{u,d}$ forces the vacuum of the potential to `lock' into the sparse form
\renewcommand*{\arraystretch}{.8}
\begin{equation}
	\langle \tilde{U}_u \rangle = \begin{pmatrix} 0 & \cos\theta_u & \sin\theta_u \\ 1 & 0 & 0 \\ 0 & -\sin\theta_u & \cos\theta_u \end{pmatrix}\,, \qquad \langle \tilde{U}_d \rangle = \begin{pmatrix} 0 & \cos\theta_d & \sin\theta_d \\ 0 & -\sin\theta_d & \cos\theta_d  \\ 1 & 0 & 0 \end{pmatrix}\,,
\end{equation}
in which the $\theta_{u,d}$ are flat directions of the potential. At this vacuum, and asserting the natural expectation $\langle r_{u,d} \rangle \sim \Lambda$, the operator $\mathcal{O}_{W',Z'}$ reduces to
\begin{equation}
	\label{eqn:QZWpI3}
	\mathcal{O}_{W',Z'} \sim  \bar{c}_R \slashed{Z}' c_R +  \bar{b}_R \slashed{Z}' b_R + \bar{c}_R \slashed{W}' b_R + \text{h.c.}\,.
\end{equation}
Equivalently, this potential enforces the flavon vacuum $\langle \lambda_u \rangle = \{0,1,0\}$ and $\langle \lambda_d \rangle = \{0,0,1\}$ in the quark mass basis, corresponding to the couplings in Eq.~\eqref{eqn:FL23}.

%%%%%%%%%%%%%%%%%%%%%%%%%%%%%
\section{Symmetry breaking beyond the minimal model.} 
%%%%%%%%%%%%%%%%%%%%%%%%%%%%%%%
\label{sec:beyondminimal}
It is possible to break the relation between $W'$ and $Z'$ masses in Eq.~\eqref{eq:masses} by introducing additional sources of $SU(2)_V \times U(1)' \to U(1)_Y$ breaking. As an example consider that in addition to $H_V$ another complex scalar, $\Phi$, obtains a vev. We take $\Phi$ to be in a $(2j+1)$-dimensional representation of $SU(2)_V$ and to carry a $U(1)'$ charge $Y'=j$. A phenomenologically viable possibility is that only
the component of the $\Phi$ multiplet that has zero hypercharge, the $\Phi_{-j}$, acquires the vacuum expectation value 
(we use the notation $T_V^3 \, \Phi_{m} = m \, \Phi^{m}$, where $m= -j,...,j$)
\beq
\vev{\Phi_{-j}} = \frac{v_{j}}{\sqrt{2}}.
\eeq
The extra contributions to the $W'$ and $Z'$ masses are then, respectively,
 \beq
 \Delta m_{W'}^2 = \frac{g_V^2 v_j^2}{2}\,j, ~~~{\rm and}~~ \Delta m_{Z'}^2 = \frac{g_V^2 v_j^2}{c_V^2}\,j^2.
 \eeq
 For large enough $j$ it is therefore possible to keep $W'$ relatively light, as dictated by the $\mathcal{R}(\dds)$ anomaly, and at the same time increase the $Z'$ mass above the experimental bounds. 
 In Section~\ref{sec:constr} we will see that a large enough splitting is obtained already for $j=1$, i.e., for $\Phi$ that is an $SU(2)_V$ triplet. We parametrize the ratio of the two vevs, $v_j$ and $v_V$,  through
 \beq
\frac{ v_j}{v_V} = \frac{a}{ \sqrt{2 j}},
 \eeq 
 so that $a$ is a continuous parameter that can take values $a\in [0, \infty)$. With this parametrization
 \beq
 \label{eq:mW':gen}
 m_W' = \frac{g_V v_V}{2} \sqrt{1+a^2},~~~{\rm and}~~~m_Z' = \frac{g_V v_V}{2 c_V} \sqrt{1+2 a^2 j}.
 \eeq
  The $Z'$ mass is arbitrarily increased in the limit of large $j$, keeping $a$ fixed, while $m_W'$ remains unchanged in that limit.

\bibliographystyle{h-physrev}
\bibliography{WprRDbiblio}

\begin{thebibliography}{10}

\bibitem{Lees:2012xj}
BaBar Collaboration, J.~P. Lees {\em et~al.},
\newblock Phys. Rev. Lett. {\bf 109}, 101802 (2012), 1205.5442.

\bibitem{Lees:2013uzd}
BaBar, J.~P. Lees {\em et~al.},
\newblock Phys. Rev. {\bf D88}, 072012 (2013), 1303.0571.

\bibitem{Huschle:2015rga}
Belle, M.~Huschle {\em et~al.},
\newblock Phys. Rev. {\bf D92}, 072014 (2015), 1507.03233.

\bibitem{Abdesselam:2016cgx}
Belle Collaboration, A.~Abdesselam {\em et~al.},
\newblock (2016), 1603.06711.

\bibitem{Abdesselam:2016xqt}
Belle Collaboration, A.~Abdesselam {\em et~al.},
\newblock (2016), 1608.06391.

\bibitem{Aaij:2015yra}
LHCb Collaboration, R.~Aaij {\em et~al.},
\newblock Phys. Rev. Lett. {\bf 115}, 111803 (2015), 1506.08614,
\newblock [Addendum: Phys. Rev. Lett. 115, no.15, 159901 (2015)].

\bibitem{HFAG}
Heavy Flavor Averaging Group, Y.~Amhis {\em et~al.},
\newblock (2016), 1612.07233,
\newblock and updates at \url{http://www.slac.stanford.edu/xorg/hfag/}.

\bibitem{Bernlochner:2017jka}
F.~U. Bernlochner, Z.~Ligeti, M.~Papucci, and D.~J. Robinson,
\newblock Phys. Rev. {\bf D95}, 115008 (2017), 1703.05330.

\bibitem{Bigi:2017jbd}
D.~Bigi, P.~Gambino, and S.~Schacht,
\newblock JHEP {\bf 11}, 061 (2017), 1707.09509.

\bibitem{Crivellin:2012ye}
A.~Crivellin, C.~Greub, and A.~Kokulu,
\newblock Phys. Rev. {\bf D86}, 054014 (2012), 1206.2634.

\bibitem{Celis:2012dk}
A.~Celis, M.~Jung, X.-Q. Li, and A.~Pich,
\newblock JHEP {\bf 01}, 054 (2013), 1210.8443.

\bibitem{Crivellin:2013wna}
A.~Crivellin, A.~Kokulu, and C.~Greub,
\newblock Phys. Rev. {\bf D87}, 094031 (2013), 1303.5877.

\bibitem{Greljo:2015mma}
A.~Greljo, G.~Isidori, and D.~Marzocca,
\newblock JHEP {\bf 07}, 142 (2015), 1506.01705.

\bibitem{Boucenna:2016qad}
S.~M. Boucenna, A.~Celis, J.~Fuentes-Martin, A.~Vicente, and J.~Virto,
\newblock JHEP {\bf 12}, 059 (2016), 1608.01349.

\bibitem{Dorsner:2016wpm}
I.~Dorsner, S.~Fajfer, A.~Greljo, J.~F. Kamenik, and N.~Kosnik,
\newblock Phys. Rept. {\bf 641}, 1 (2016), 1603.04993.

\bibitem{Bauer:2015knc}
M.~Bauer and M.~Neubert,
\newblock Phys. Rev. Lett. {\bf 116}, 141802 (2016), 1511.01900.

\bibitem{Fajfer:2015ycq}
S.~Fajfer and N.~Kosnik,
\newblock Phys. Lett. {\bf B755}, 270 (2016), 1511.06024.

\bibitem{Barbieri:2015yvd}
R.~Barbieri, G.~Isidori, A.~Pattori, and F.~Senia,
\newblock Eur. Phys. J. {\bf C76}, 67 (2016), 1512.01560.

\bibitem{Becirevic:2016yqi}
D.~Becirevic, S.~Fajfer, N.~Kosnik, and O.~Sumensari,
\newblock Phys. Rev. {\bf D94}, 115021 (2016), 1608.08501.

\bibitem{Hiller:2016kry}
G.~Hiller, D.~Loose, and K.~Schönwald,
\newblock JHEP {\bf 12}, 027 (2016), 1609.08895.

\bibitem{Crivellin:2017zlb}
A.~Crivellin, D.~Müller, and T.~Ota,
\newblock JHEP {\bf 09}, 040 (2017), 1703.09226.

\bibitem{Fajfer:2012jt}
S.~Fajfer, J.~F. Kamenik, I.~Nisandzic, and J.~Zupan,
\newblock Phys. Rev. Lett. {\bf 109}, 161801 (2012), 1206.1872.

\bibitem{Freytsis:2015qca}
M.~Freytsis, Z.~Ligeti, and J.~T. Ruderman,
\newblock Phys. Rev. {\bf D92}, 054018 (2015), 1506.08896.

\bibitem{Bardhan:2016uhr}
D.~Bardhan, P.~Byakti, and D.~Ghosh,
\newblock JHEP {\bf 01}, 125 (2017), 1610.03038.

\bibitem{Li:2016vvp}
X.-Q. Li, Y.-D. Yang, and X.~Zhang,
\newblock JHEP {\bf 08}, 054 (2016), 1605.09308.

\bibitem{Alonso:2016oyd}
R.~Alonso, B.~Grinstein, and J.~Martin~Camalich,
\newblock Phys. Rev. Lett. {\bf 118}, 081802 (2017), 1611.06676.

\bibitem{Celis:2016azn}
A.~Celis, M.~Jung, X.-Q. Li, and A.~Pich,
\newblock Phys. Lett. {\bf B771}, 168 (2017), 1612.07757.

\bibitem{Faroughy:2016osc}
D.~A. Faroughy, A.~Greljo, and J.~F. Kamenik,
\newblock Phys. Lett. {\bf B764}, 126 (2017), 1609.07138.

\bibitem{Feruglio:2016gvd}
F.~Feruglio, P.~Paradisi, and A.~Pattori,
\newblock Phys. Rev. Lett. {\bf 118}, 011801 (2017), 1606.00524.

\bibitem{Feruglio:2017rjo}
F.~Feruglio, P.~Paradisi, and A.~Pattori,
\newblock JHEP {\bf 09}, 061 (2017), 1705.00929.

\bibitem{Buttazzo:2017ixm}
D.~Buttazzo, A.~Greljo, G.~Isidori, and D.~Marzocca,
\newblock JHEP {\bf 11}, 044 (2017), 1706.07808.

\bibitem{DiLuzio:2017vat}
L.~Di~Luzio, A.~Greljo, and M.~Nardecchia,
\newblock Phys. Rev. {\bf D96}, 115011 (2017), 1708.08450.

\bibitem{Bordone:2017bld}
M.~Bordone, C.~Cornella, J.~Fuentes-Martin, and G.~Isidori,
\newblock Phys. Lett. {\bf B779}, 317 (2018), 1712.01368.

\bibitem{Barbieri:2017tuq}
R.~Barbieri and A.~Tesi,
\newblock Eur. Phys. J. {\bf C78}, 193 (2018), 1712.06844.

\bibitem{Blanke:2018sro}
M.~Blanke and A.~Crivellin,
\newblock (2018), 1801.07256.

\bibitem{Greljo:2018tuh}
A.~Greljo and B.~A. Stefanek,
\newblock (2018), 1802.04274.

\bibitem{Marzocca:2018wcf}
D.~Marzocca,
\newblock (2018), 1803.10972.

\bibitem{Robinson:2018EFT}
D.~J. Robinson, B.~Shakya, and J.~Zupan,
\newblock (to appear).

\bibitem{Cvetic:2017gkt}
G.~Cvetic, F.~Halzen, C.~S. Kim, and S.~Oh,
\newblock Chin. Phys. {\bf C41}, 113102 (2017), 1702.04335.

\bibitem{Ligeti:2016npd}
Z.~Ligeti, M.~Papucci, and D.~J. Robinson,
\newblock JHEP {\bf 01}, 083 (2017), 1610.02045.

\bibitem{Abdesselam:2017kjf}
Belle Collaboration, A.~Abdesselam {\em et~al.},
\newblock (2017), 1702.01521.

\bibitem{Bailey:2014tva}
Fermilab Lattice, MILC, J.~A. Bailey {\em et~al.},
\newblock Phys. Rev. {\bf D89}, 114504 (2014), 1403.0635.

\bibitem{Lattice:2015rga}
Fermilab Lattice and MILC Collaborations, J.~A. Bailey {\em et~al.},
\newblock Phys. Rev. {\bf D92}, 034506 (2015), 1503.07237.

\bibitem{Colquhoun:2015oha}
HPQCD, B.~Colquhoun {\em et~al.},
\newblock Phys. Rev. {\bf D91}, 114509 (2015), 1503.05762.

\bibitem{PDG}
Particle Data Group, C.~Patrignani {\em et~al.},
\newblock Chin. Phys. {\bf C40}, 100001 (2016).

\bibitem{Hammer_paper}
F.~Bernlochner, S.~Duell, Z.~Ligeti, M.~Papucci, and D.~J. Robinson,
\newblock In preparation  (2018).

\bibitem{Fajfer:2013wca}
S.~Fajfer, A.~Greljo, J.~F. Kamenik, and I.~Mustac,
\newblock JHEP {\bf 07}, 155 (2013), 1304.4219.

\bibitem{Aaboud:2017sjh}
ATLAS, M.~Aaboud {\em et~al.},
\newblock JHEP {\bf 01}, 055 (2018), 1709.07242.

\bibitem{Aaboud:2017buh}
ATLAS, M.~Aaboud {\em et~al.},
\newblock JHEP {\bf 10}, 182 (2017), 1707.02424.

\bibitem{Aaboud:2018vgh}
ATLAS, M.~Aaboud {\em et~al.},
\newblock (2018), 1801.06992.

\bibitem{CMS:2016ppa}
CMS,
\newblock CERN Report No. CMS-PAS-EXO-16-006, 2016 (unpublished).

\bibitem{Sirunyan:2017lzl}
CMS, A.~M. Sirunyan {\em et~al.},
\newblock (2017), 1708.02510.

\bibitem{Leurer:1992wg}
M.~Leurer, Y.~Nir, and N.~Seiberg,
\newblock Nucl. Phys. {\bf B398}, 319 (1993), hep-ph/9212278.

\bibitem{Leurer:1993gy}
M.~Leurer, Y.~Nir, and N.~Seiberg,
\newblock Nucl. Phys. {\bf B420}, 468 (1994), hep-ph/9310320.

\bibitem{Patrignani:2016xqp}
Particle Data Group, C.~Patrignani {\em et~al.},
\newblock Chin. Phys. {\bf C40}, 100001 (2016).

\bibitem{Bona:2007vi}
UTfit, M.~Bona {\em et~al.},
\newblock JHEP {\bf 03}, 049 (2008), 0707.0636.

\bibitem{Bevan:2014cya}
A.~Bevan {\em et~al.},
\newblock (2014), 1411.7233.

\bibitem{Martinelli:2015}
G.~Martinelli,
\newblock Unitarity triangle fits: Standard model \&search for new physics.

\bibitem{Babu:1988yq}
K.~S. Babu and X.~G. He,
\newblock Mod. Phys. Lett. {\bf A4}, 61 (1989).

\bibitem{Balakrishna:1988bn}
B.~S. Balakrishna and R.~N. Mohapatra,
\newblock Phys. Lett. {\bf B216}, 349 (1989).

\bibitem{Borah:2017leo}
D.~Borah and A.~Dasgupta,
\newblock JCAP {\bf 1706}, 003 (2017), 1702.02877.

\bibitem{Lavoura:2003xp}
L.~Lavoura,
\newblock Eur. Phys. J. {\bf C29}, 191 (2003), hep-ph/0302221.

\bibitem{Wong:1992qa}
G.-G. Wong,
\newblock Phys. Rev. {\bf D46}, 3987 (1992).

\bibitem{Bezrukov:2009th}
F.~Bezrukov, H.~Hettmansperger, and M.~Lindner,
\newblock Phys. Rev. {\bf D81}, 085032 (2010), 0912.4415.

\bibitem{Scherrer:1984fd}
R.~J. Scherrer and M.~S. Turner,
\newblock Phys. Rev. {\bf D31}, 681 (1985).

\bibitem{Asaka:2006ek}
T.~Asaka, M.~Shaposhnikov, and A.~Kusenko,
\newblock Phys. Lett. {\bf B638}, 401 (2006), hep-ph/0602150.

\bibitem{Essig:2013goa}
R.~Essig, E.~Kuflik, S.~D. McDermott, T.~Volansky, and K.~M. Zurek,
\newblock JHEP {\bf 11}, 193 (2013), 1309.4091.

\bibitem{Anelli:2015pba}
SHiP, M.~Anelli {\em et~al.},
\newblock (2015), 1504.04956.

\bibitem{Chou:2016lxi}
J.~P. Chou, D.~Curtin, and H.~J. Lubatti,
\newblock Phys. Lett. {\bf B767}, 29 (2017), 1606.06298.

\bibitem{Feng:2017uoz}
J.~Feng, I.~Galon, F.~Kling, and S.~Trojanowski,
\newblock Phys. Rev. {\bf D97}, 035001 (2018), 1708.09389.

\bibitem{Gligorov:2017nwh}
V.~V. Gligorov, S.~Knapen, M.~Papucci, and D.~J. Robinson,
\newblock Phys. Rev. {\bf D97}, 015023 (2018), 1708.09395.

\bibitem{Knapen:2015hia}
S.~Knapen and D.~J. Robinson,
\newblock Phys. Rev. Lett. {\bf 115}, 161803 (2015), 1507.00009.

\bibitem{Altmannshofer:2017uvs}
W.~Altmannshofer, S.~Gori, D.~J. Robinson, and D.~Tuckler,
\newblock JHEP {\bf 03}, 129 (2018), 1712.01847.

\end{thebibliography}

\end{document}